\documentclass[prodmode,acmcsur]{acmsmall} % Aptara syntax
\newcommand{\revision}[1]{\textcolor{red}{\textbf{#1}}}
\renewcommand{\revision}[1]{{#1}}

\usepackage[utf8]{inputenc}
\usepackage[T1]{fontenc}
\usepackage[greek,english]{babel}
\usepackage{graphicx}
\usepackage{tikz}
\graphicspath{ {images/} }
\usepackage{amsmath}
\usepackage{amssymb}
\usepackage{mwe}
\usetikzlibrary{matrix,shapes,arrows,positioning,chains,calc}
\usepackage{array}
\newcolumntype{L}{>{\centering\arraybackslash}m{3cm}}
\usepackage{siunitx}
\usepackage{multirow}
\usepackage{array,longtable,booktabs}
\usepackage{caption}
\usepackage{comment}

\usepackage{tablefootnote}

\usepackage{pifont}% http://ctan.org/pkg/pifont
\newcommand{\cmark}{\ding{51}}%
\newcolumntype{C}[1]{>{\centering\let\newline\\\arraybackslash\hspace{0pt}}m{#1}}

\newcommand{\specialcell}[2][c]{%
  \begin{tabular}[#1]{@{}c@{}}#2\end{tabular}}

\newcolumntype{K}[1]{>{\centering\arraybackslash}p{#1}}

\begin{document}

% Page heads
\markboth{A. Acar et al.}{A Survey on Homomorphic Encryption Schemes:\\ Theory and Implementation}

% Title portion
\title{A Survey on Homomorphic Encryption Schemes:\\ Theory and Implementation}
\author{ABBAS ACAR, HIDAYET AKSU, and A. SELCUK ULUAGAC
\affil{Florida International University} 
MAURO CONTI
\affil{University of Padua}}
% NOTE! Affiliations placed here should be for the institution where the
%       BULK of the research was done. If the author has gone to a new
%       institution, before publication, the (above) affiliation should NOT be changed.
%       The authors 'current' address may be given in the "Author's addresses:" block (below).
%       So for example, Mr. Abdelzaher, the bulk of the research was done at UIUC, and he is
%       currently affiliated with NASA.

\begin{abstract}
Legacy encryption systems depend on sharing a key (public or private) among the peers involved in exchanging  an encrypted message. However, this approach poses privacy concerns. The users or service providers with the key have exclusive rights on the data. Especially with popular cloud services, the control over the privacy of the sensitive data is lost. Even when the keys are not shared, the encrypted material is shared with a third party that does not necessarily need to access the content. Moreover, untrusted servers, providers, and cloud operators can keep identifying elements of users long after users end the relationship with the services. Indeed, \textit{Homomorphic Encryption} (HE), a special kind of encryption scheme, can address these concerns as it allows any third party to operate on the encrypted data without decrypting it in advance. Although this extremely useful feature of the HE scheme has been known for over 30 years, the first plausible and achievable \textit{Fully Homomorphic Encryption} (FHE) scheme, which allows any computable function to perform on the encrypted data, was introduced by Craig Gentry in 2009. Even though this was a major achievement, different implementations so far
demonstrated that FHE still needs to be improved significantly to be practical on every platform. Therefore, this survey focuses
on HE and FHE schemes. First, we present the basics of HE and the details of the well-known Partially Homomorphic Encryption
(PHE) and Somewhat Homomorphic Encryption (SWHE), which are important pillars of achieving FHE. Then, the main FHE families, which have become the base for the other follow-up
FHE schemes are presented. Furthermore, the implementations and 
\revision{recent} %new 
improvements in Gentry-type FHE schemes are also surveyed. Finally, further research directions are discussed. 
% We believe this survey can 
This survey \revision{is intended to} 
give a clear knowledge and foundation to researchers and practitioners interested in knowing, applying, as well as extending the state of the art HE, PHE, SWHE, and FHE systems.
\end{abstract}



\acmformat{Abbas Acar, Hidayet Aksu, A. Selcuk Uluagac, and Mauro Conti, 2016. A survey on homomorphic encryption schemes: theory and implementation.}
% At a minimum you need to supply the author names, year and a title.
% IMPORTANT:
% Full first names whenever they are known, surname last, followed by a period.
% In the case of two authors, 'and' is placed between them.
% In the case of three or more authors, the serial comma is used, that is, all author names
% except the last one but including the penultimate author's name are followed by a comma,
% and then 'and' is placed before the final author's name.
% If only first and middle initials are known, then each initial
% is followed by a period and they are separated by a space.
% The remaining information (journal title, volume, article number, date, etc.) is 'auto-generated'.

\begin{bottomstuff}
Author's addresses: A. Acar, H. Aksu,  {and} A. S. Uluagac,
Electrical and Computer Engineering, Florida International University, Miami, FL-33199; emails: {aacar001,haksu,suluagac}@fiu.edu; M. Conti, Department of Mathematics, University of Padua, Padua, Italy and email: conti@math.unipd.it. 
\end{bottomstuff}

\maketitle

\textbf{This paper is an early draft of the survey that is being submitted to ACM CSUR and has been uploaded to arXiv for feedback from stakeholders.}

\section{Introduction}

\par In ancient Greeks, the term "\textgreek{ὁμός}" (homos) was used in the meaning of "same" while "\textgreek{μορφή}" (morphe) was used for "shape" \cite{liddell1896intermediate}. Then, the term \textit{homomorphism} is coined and used in different areas. In abstract algebra, homomorphism is defined as a map preserving all the algebraic structures between the domain and range of an algebraic set \cite{malik2007mth}. The map is simply a function, i.e., an operation, which takes the inputs from the set of domain and outputs an element in the range, (e.g., addition, multiplication). In the cryptography field, the homomorphism is used as an encryption type. The \textit{Homomorphic Encryption} (HE) is a kind of encryption scheme which allows a third party (e.g., cloud, service provider) to perform certain computable functions on the encrypted data while preserving the features of the function and format of the encrypted data. Indeed, this homomorphic encryption corresponds to a mapping in the abstract algebra. As an example for an additively HE scheme, for sample messages $m_1$ and $m_2$, one can obtain $E(m_1+m_2)$ by using $E(m_1)$ and $E(m_2)$ without knowing $m_1$ and $m_2$ explicitly, where $E$ denotes the encryption function.
\par Normally, encryption is a crucial mechanism to preserve the privacy of any sensitive information. However, the conventional encryption schemes can not work on the encrypted data without decrypting it first. In other words, the users have to sacrifice their privacy to make use of cloud services such as file storing, sharing and collaboration. Moreover, untrusted servers, providers, popular cloud operators can keep physically identifying elements of users long after users end the relationship with the services \cite{AppleSiriFact}. This is a major privacy concern for users. In fact, it would be perfect if there existed a scheme which would not restrict the operations to be computed on the encrypted data while it would be still encrypted. From a historical perspective in cryptology, the term \textit{homomorphism} is used for the first time by Rivest, Adleman, and Dertouzous \cite{rivest1978data} in 1978 as a possible solution to the computing without decrypting problem. This given basis in \cite{rivest1978data} has led to numerous attempts by researchers around the world to design such a homomorphic scheme with a large set of operations. In this work, the primary motivation is to survey the HE schemes focusing on the most recent improvements in this field, including \textit{partially}, \textit{somewhat}, and \textit{fully} HE schemes.

\par
\begin{figure}[t] 
\centering
\resizebox{!}{3.5cm}{  %8.65cm
\framebox{
\begin{tikzpicture}[>=stealth,  controller/.style={draw, rectangle, minimum height=3em, minimum width=6em}]
\node[ label=above:{ {\small C: $m$, $Enc$, $Dec$, $f()$}}] (A) {\includegraphics[height=3.5cm]{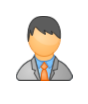}};
\node[right=4cm of A, label=above:{{\small S: $Eval$}}] (B) {\includegraphics[height=3.5cm]{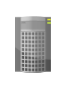}};

\node [controller, right = of B, xshift=-1cm, yshift=0.1cm, text centered, text width=1.8cm] (control1) {{\small \framebox{4} S evalutes $f()$ homomorphically}}; 

\node [controller, left = of A, yshift=1cm, xshift=1cm, text centered, text width=4cm] (control2) {{\small \framebox{1} C encrypts his message $m$, $Enc(m)$}};
\node [controller, left = of A, yshift=-1cm, xshift=1cm, text centered, text width=4cm] (control3) {{\small \framebox{6} C computes $Dec(Enc(f(m)))=f(m)$, and recovers $f(m)$}};  

\draw[->,thick] (A.30) -- node[above]{ {\small \framebox{2} C sends $Enc(m)$ to store }} (B.145);
\draw[shorten <=0cm,shorten >=2.2cm, -latex, thick] (A.0) -- node[above left=0cm]{{\small \framebox{3} C queries, $f()$}} (B.0);

\draw[<-,thick] (A.-40) -- node[above]{{\small \framebox{5} S returns $Enc(f(m))$}} (B.-135);
\end{tikzpicture}
}
}
%\vspace{7.5pt}
\caption{A simple client-server HE scenario, where C is Client and S is Server \label{basic} }
\vspace{-10pt}
\end{figure}
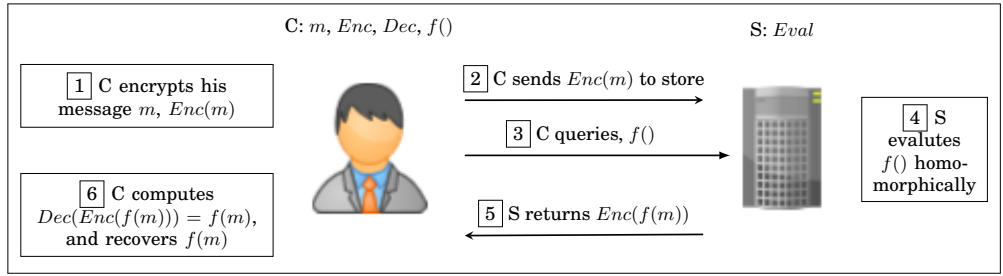

 A simple motivational HE example for a sample cloud application is illustrated in Figure~\ref{basic}. In this scenario, the client, $C$, first encrypts her private data (Step 1), then sends the encrypted data to the cloud servers, $S$, (Step 2). When the client wants to perform a function \revision{(i.e., query)}, $f()$, over her own data, she sends the function to \revision{the} server (Step 3). The server performs a homomorphic operation over the encrypted data using the $Eval$ function, i.e., computes $f()$ blindfolded (Step 4) and returns the encrypted result to the client (Step 5). Finally, the client recovers the data with her own secret key and obtains $f(m)$ (Step 6). As seen in this simple example, the homomorphic operation, $Eval()$, at the server side does not require the private key of the client and allows various operations such as addition and multiplication on the encrypted client data.
\par An early attempt to compute functions/operations on encrypted data is Yao's \textit{garbled circuit}\footnote{A circuit is the set of connected gates (e.g., AND and XOR gates in boolean circuits), where the evaluation is completed by calculating the output of each gate in turn.} study \cite{yao1982protocols}. Yao proposed two party communication protocol as a solution to the Millionaires' problem, which compares the wealth of two rich people without revealing the exact amount to each other. However, in Yao's \textit{garbled circuit} solution, ciphertext size grows at least linearly with the computation of every gate in the circuit. This yields a very poor efficiency in terms of computational overhead and too much complexity in its communication protocol. Until Gentry's breakthrough in~\cite{gentry2009fully}, all the attempts \cite{rivest1978method,goldwasser1982probabilistic,elgamal1985public,benaloh1994dense,naccache1998new,okamoto1998new,paillier1999public,damgaard2001generalisation,kawachi2007multi,yao1982protocols,boneh2005evaluating,814630,ishai2007evaluating} have allowed either one type of operation or limited number of operations on the encrypted data. Moreover, some of the attempts are even limited over a specific type of set (e.g., branching programs). In fact, all these different HE attempts can neatly be categorized under three types of schemes with respect to the number of allowed operations on the encrypted data as follows: (1) \emph{Partially Homomorphic Encryption} (PHE) allows only one type of operation with an unlimited number of times (i.e., no bound on the number of usages). (2) \emph{Somewhat Homomorphic Encryption} (SWHE) allows some types of operations with a limited number of times. (3) \emph{Fully Homomorphic Encryption} (FHE) allows an unlimited number of operations with unlimited number of times.

%After Yao's work \cite{yao1982protocols}, even the existence of the homomorphic scheme suggested in \cite{rivest1978data} was questioned.

\par PHE schemes are deployed in some applications like e-voting \cite{benaloh1987verifiable} or Private Information Retrieval (PIR) \cite{kushilevitz1997replication}. However, these applications were restricted in terms of the types of homomorphic evaluation operations. In other words, PHE schemes can only be used for particular applications, whose algorithms include only addition or multiplication operation. On the other hand, the SWHE schemes support both addition and multiplication. Nonetheless, in SWHE schemes that are proposed before the first FHE scheme, the size of the ciphertexts grows with each homomorphic operation and hence the maximum number of allowed homomorphic operations is limited. These issues put a limit on the use of PHE and SWHE schemes in real-life applications. Eventually, the increasing popularity of cloud-based services accelerated the design of HE schemes which can support an  arbitrary number of homomorphic operations with random functions, i.e. FHE. Gentry's FHE scheme is the first plausible and achievable FHE scheme \cite{gentry2009fully}. It is based on ideal-lattices in math and it is not only a description of the scheme, but also a powerful framework for achieving FHE. However, it is conceptually and practically not a realistic scheme. Especially, the \textit{bootstrapping} part, which is the intermediate refreshing procedure of a processed ciphertext, is too costly in terms of computation. Therefore, a lot of follow-up improvements and new schemes were proposed in the following years. 

\par \textit{Contribution:} In \revision{this} work, we provide a comprehensive survey of all the main FHE schemes as of this writing. We also \revision{cover} a survey of important PHE and SWHE schemes as they are the first works in accomplishing the FHE dream and are still popular as FHE schemes are very costly. Furthermore, we \revision{include} the FHE implementations focusing on the improvements with each scheme. FHE attracts the interest of people from very different research areas in terms of theoretical, implementation, and application perspectives. \revision{This survey is structured} to provide an easy digest of the relatively \revision{complex} homomorphic encryption topic. For instance, while a mathematician focuses on the improvement in theoretical perspective, a hardware designer tries to improve the efficiency of FHE by implementing on GPU instead of CPU. All such different attempts make it harder to follow recent works. \revision{Therefore}, it is important to collect and categorize the existing FHE works focusing on recent improvements. In addition, \revision{we mention} the challenges and future perspectives of HE to motivate the researchers and practitioners to explore and improve the performance of HE schemes and their applications. This survey \revision{is intended to} give a clear knowledge foundation to researchers and practitioners interested in knowing, applying, as well as extending state of the art HE systems.

%\{contribution} \par With this survey, we aim to fill the gap between the researchers and practitioners and provide an in-depth insight about the state of the art HE schemes. For this purpose, this survey provides an understanding of the basics of HE theory, the notable PHE, SWHE, FHE schemes and the improvements of FHE schemes. Furthermore, it also provides the implementations of the proposed schemes and primitively introduced applications. 

\par \textit{Organization:} The reminder of the paper is organized as follows: In Section~\ref{sec:hom_enc}, descriptions of different HE schemes, PHE, SWHE, and FHE schemes are presented. Then, in Section~\ref{sec:implementations}, different implementations of SWHE and FHE schemes, which were introduced after Gentry's work, are given and their performances are discussed. Finally, in Section~\ref{sec:further}, further research directions and lessons learned are given and the paper is concluded.

%%%%%%%%%%%%%%%%%%%%%%%%%%%%%%%%%%%%%%%%%%%%%%%%%%%%%%%%%%%%%%%%%%%%%%%

\section{Related Work}
\par Like our work in this paper, there are similar useful surveys in the literature. In fact, unfortunately, some of the surveys only cover the theoretical information of the schemes as in~\cite{parmar2014survey,ahilastate} and some of them are directly for expert readers and mathematicians as in~\cite{vaikuntanathan2011computing,silverberg2013fully,gentry2014computing}. Compared to these surveys, our \revision{survey} has a broad reader perspective including researchers and practitioners interested in the advances and implementations in the field of HE, especially FHE. Furthermore, while the survey in~\cite{aguilar2013recent} only covers the signal processing applications, other in \cite{hrestak2014homomorphic} covers a few FHEs on only cloud applications. Since our survey is not limited to specific application \revision{areas}, we do not articulate these specific application areas in detail but we list the theory and implementation of all existing HE schemes, which can be used in possible futuristic application areas with recent advancements. After \cite{fontaine2007survey} and  \cite{akinwande2009advances}, many HE schemes were introduced. Compared to these useful surveys, our survey focuses on the most recent HE schemes, since most of the significant improvements are introduced recently (after 2009). Although~\cite{moore2014practical} is one of the most recent surveys, it focuses on the hardware implementation solutions of FHE schemes. This survey is not limited to hardware solutions, as, in addition to hardware solutions, it covers software solutions of implementations as well in the implementation section. After~\cite{sen2013homomorphic,wu2015fully}, several new FHE schemes, which improves FHE in a sufficiently great way as to be worthy of attention, were proposed in the literature. \revision{Finally, it is worth mentioning that \cite{armknecht2015guide} provides a systematic explanation of the new terminology related to FHE, where they do not present the HE schemes and implementations in detail. Compared to these useful prior surveys, nonetheless, our survey is intrinsically different from the aforementioned surveys.}
%we hold as an opinion that there are adequate differences to make this survey recognizably different in nature from aforementioned surveys.

%%%%%%%%%%%%%%%%%%%%%%%%%%%%%%%%%%%%%%%%%%%%%%%%%%%%%%%%%%%%

\section{Homomorphic Encryption Schemes} \label{sec:hom_enc}
\par In this section, we explain the basics of HE theory. Then, we present notable PHE, SWHE and FHE schemes. For each scheme, we also give a brief description of the scheme. 
 
\newtheorem{mydef}{Definition}
\begin{mydef}
An encryption scheme is called \textit{homomorphic} over an operation '$\star$'  if it supports the following equation:
\begin{equation}
E(m_1) \star E(m_2)=E(m_1 \star m_2), \ \ \forall m_1,m_2 \in M,
\end{equation}
 where $E$ is the encryption algorithm and M is the set of all possible messages.
\end{mydef}
\par In order to create an encryption scheme allowing the homomorphic evaluation of arbitrary function, it is sufficient to allow only addition and multiplication operations because addition and multiplication are functionally complete sets over finite sets. Particularly, any boolean circuit can be represented using  only XOR (addition) and AND (multiplication) gates. While an HE scheme can use the same key for both encryption and decryption (symmetric), it can also be designed to use the different keys to encrypt and decrypt (asymmetric). A generic method to transform symmetric and asymmetric HE schemes to each other is demonstrated in~\cite{rothblum2011homomorphic}.

An HE scheme is primarily characterized by four operations: $KeyGen$, $Enc$, $Dec$, and $Eval$. $KeyGen$ is the operation, which generates a secret and public key pair for the asymmetric version of HE or a single key for the symmetric version. Actually, $KeyGen$, $Enc$ and $Dec$ are not different from their classical tasks in conventional encryption schemes. However, $Eval$ is an HE-specific operation, which takes ciphertexts as input and outputs a ciphertext corresponding to a functioned plaintext. $Eval$ performs the function $f()$ over the ciphertexts $(c_1,c_2)$ without seeing the messages $(m_1,m_2)$. $Eval$ takes ciphertexts as input and outputs evaluated ciphertexts. The most crucial point in this homomorphic encryption is that the format of the ciphertexts after an evaluation process must be preserved in order to be decrypted correctly. In addition, the size of the ciphertext should also be constant to support unlimited number of operations. Otherwise, the increase in the ciphertext size will require more resources and this will limit the number of operations. 

%The commutative diagram in Figure~\ref{diagram} depicts the relationship among the four major operations. The diagram~\cite{gentry2014computing} is simplified to show only one homomorphic encryption with two ciphertexts:

\begin{figure*}[t]
  \centering
  \framebox{
  \centering
    \includegraphics[clip, trim=0.3cm 6cm 0.5cm 8cm, width=\textwidth]{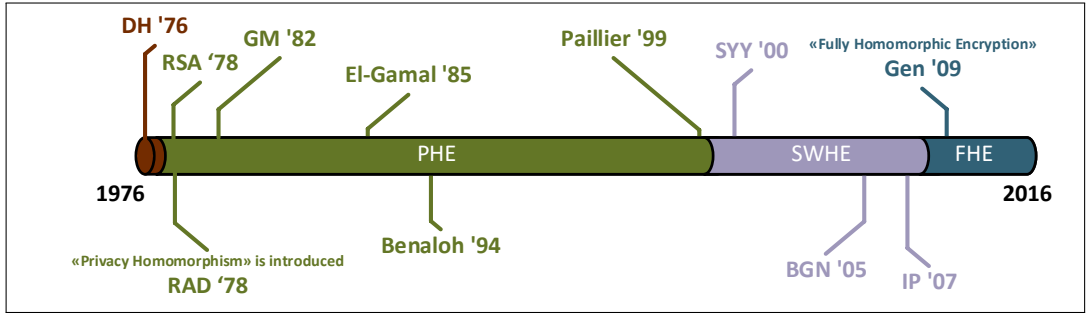}
    }
      \caption{Timeline of HE schemes until Gentry's first FHE scheme}
      \label{timeline}
\end{figure*}

%\par Moreover, HE is computationally heavy and very complicated. There are different HE schemes in the literature to overcome this.
Of all \revision{HE} schemes in the literature, PHE schemes support $Eval$ function for only either addition or multiplication, SWHE schemes support for only limited number of operations or some limited circuits (e.g., branching programs) while FHE schemes supports the evaluation of arbitrary functions (e.g., searching, sorting, max, min, etc.) with unlimited number of times over ciphertexts. The well-known PHE, SWHE, and FHE schemes are summarized in the timeline in Figure~\ref{timeline} and are explained in the following sections with a greater detail. \revision{The interest in the area of HE significantly increased after the work of Gentry~\cite{gentry2009fully} in 2009. Therefore, we articulate the HE schemes, FHE anymore, after Gentry's work in a greater detail and we also discuss their implementations and recent techniques to make it faster in Section~\ref{sec:implementations}. Here, we start with the PHE schemes, which are the first stepping stones for FHE schemes.}

\subsection{Partially Homomorphic Encryption Schemes}
%In this section, we articulate the details of PHE schemes.
There are several useful
PHE examples~\cite{rivest1978method,goldwasser1982probabilistic,elgamal1985public,benaloh1994dense,naccache1998new,okamoto1998new,paillier1999public,damgaard2001generalisation,kawachi2007multi} in the literature. Each has improved the PHE in some way. However, in this section, we primarily focus on major PHE schemes that are the basis for many other PHE schemes.
\subsubsection{RSA}
RSA is an early example of PHE and introduced by Rivest, Shamir, and Adleman~\cite{rivest1978method} shortly after the invention of public key cryptography by Diffie Helman~\cite{diffie1976new}. RSA is the first feasible achievement of the public key cryptosystem. Moreover, the homomorphic property of RSA was shown by Rivest, Adleman, and Dertouzous~\cite{rivest1978data} just after the seminal work of RSA. Indeed, the first attested use of the term "privacy homomorphism" is introduced in~\cite{rivest1978data}. The security of the RSA cryptosystem is based on the hardness of \textit{factoring problem} of the product of two large prime numbers~\revision{\cite{montgomery1994survey}}\footnote{\revision{Here, we do not mean that RSA is secure. We mean the most basic attack on RSA (e.g., key recovering attack) has to solve the problem of factoring of two large primes. For example, plain RSA is not secure against \textit{Chosen Plaintext Attacks} (CPA) as its encryption algorithm is deterministic. We use the same idea for the rest of the paper as well. Because of the limited space, we do not discuss the details of the security of each encryption scheme.}} RSA is defined as follows:
\begin{itemize}
\item \textit{KeyGen Algorithm:} First, for large primes $p$ and $q$, $n=pq$ and $\phi=(p-1)(q-1)$ are computed. Then, $e$ is chosen such that $gcd(e,\phi)$ and $d$ is calculated by computing the multiplicative inverse of $e$ (i.e, $ed \equiv 1 \mod \phi$). Finally, $(e,n)$ is released as the public key pair while $(d,n)$ is kept as the secret key pair. 
\item \textit{Encryption Algorithm:}
First, the message is converted into a plaintext $m$ such that $0 \leq m < n$, then the RSA encryption algorithm is as follows:
\begin{equation}
c=E(m) = m^e \pmod n , \quad \forall m \in M,
\end{equation}
where $c$ is the ciphertext.
\item \textit{Decryption Algorithm:} The message $m$ can be recovered from the ciphertext $c$ using the secret key pair $(d,n)$ as follows:
\begin{equation}
m=D(c)=c^d \pmod n
\end{equation}

\item \textit{Homomorphic Property:}
For $m_1,m_2 \in M$,
\begin{equation}
E(m_1)*E(m_2)  =( m_1^e \pmod n )*( m_2^e \pmod n ) = (m_1*m_2)^e \pmod n = E(m_1*m_2).
\end{equation}
\end{itemize}
\par The homomorphic property of RSA shows that $E(m_1*m_2)$ can be directly evaluated by using $E(m_1)$ and $E(m_2)$ without decrypting them. In other words, RSA is only homomorphic over multiplication. Hence, it does not allow the homomorphic addition of ciphertexts.

\subsubsection{Goldwasser-Micali}
GM proposed the first probabilistic public key encryption scheme proposed in~\cite{goldwasser1982probabilistic}. The GM cryptosystem is based on the hardness of \textit{quadratic residuosity problem}~\revision{\cite{Kaliski2005}}. Number $a$ is called \textit{quadratic residue modulo $n$} if there exists an integer $x$ such that $x^2 \equiv a \pmod n$. Quadratic residuosity problem decides whether a given number $q$ is quadratic modulo $n$ or not.
GM cryptosystem is described as follows:
\begin{itemize}
\item \textit{KeyGen Algorithm:} Similar to RSA, $n=pq$ is computed where $p$ and $q$ are distinct large primes and then, $x$ is chosen as one of the quadratic nonresidue modulo $n$ values with $(\frac{x}{n})=1$. Finally, $(x,n)$ is published as the public key while $(p,q)$ is kept as the secret key.
\item \textit{Encryption Algorithm:} Firstly, the message ($m$) is converted into a string of bits. Then, for every bit of the message $m_i$, a quadratic nonresidue value $y_i$ is produced such that $gcd(y_i,n)=1$. Then, each bit is encrypted to $c_i$ as follows:
\begin{equation}
c_i=E(m_i)=y_i^2 x^{m_i} \pmod n, \quad \forall m_i=\{0,1\},
\end{equation}
where $m=m_0m_1...m_r$, $c=c_0c_1...c_r$ and $r$ is the block size used for the message space and  $x$ is picked from ${Z_n}^*$ at random for every encryption, where ${\mathbb{Z}_n}^*$ is the multiplicative subgroup of integers modulo $n$ which includes all the numbers smaller than $r$ and relatively prime to $r$.
\item \textit{Decryption Algorithm:} Since $x$ is picked from the set  ${Z_n}^*$ ($1<x \leq n-1$), $x$ is quadratic residue modulo $n$ for only $m_i=0$. Hence, to decrypt the ciphertext $c_i$, one decides whether $c_i$ is a quadratic residue modulo n or not; if so, $m_i$ returns $0$, else $m_i$ returns $1$.
\item \textit{Homomorphic Property:}
For each bit $m_i \in \{0,1\}$,
\begin{equation}
\begin{split}
E(m_1)*E(m_2) & =({y_1^2}{x^{m_1}} \pmod n)*({y_2^2}{x^{m_2}} \pmod n)\\ & =(y_1*y_2)^2x^{m_1 + m_2}  \pmod n =E(m_1 + m_2).
\end{split} 
\end{equation} 
\end{itemize}
\par The homomorphic property of the GM cryptosystem shows that encryption of the sum $E(m_1 \oplus m_2)$\  can be directly calculated from the separately encrypted bits, $E(m_1)$ and $E(m_2)$. Since the message and ciphertext are the elements of the set $\{0,1\}$, the operation is the same with exclusive-OR (XOR)\footnote{XOR can be thought as binary addition.} Hence, GM is homomorphic over only addition for binary numbers.
\subsubsection{\revision{El-Gamal}} 
In 1985, Taher Elgamal proposed a new public key encryption scheme~\cite{elgamal1985public} which is the improved version of the original Diffie-Hellman Key Exchange~\cite{diffie1976new} algorithm, which is based on the hardness of certain problems in discrete logarithm~\revision{\cite{kevin1990discrete}}. It is mostly used in hybrid encryption systems to encrypt the secret key of a symmetric encryption system. The \revision{El-Gamal} cryptosystem is defined as follows:
\begin{itemize}
\item \textit{KeyGen Algorithm:} A cyclic group G with order $n$ using generator $g$ is produced. In a cyclic group, it is possible to generate all the elements of the group using the powers of one of its own element. Then, $h=g^y$ computed for randomly chosen $y \in {\mathbb{Z}_n}^*$. Finally, the public key is $(G,n,g,h)$ and $x$ is the secret key of the scheme.
\item \textit{Encryption Algorithm:} The message $m$ is encrypted using $g$ and $x$, where $x$ is randomly chosen from the set $\{1,2,...,n-1\}$ and the output of the encryption algorithm is a ciphertext pair ($c=(c_1,c_2)$):
\begin{equation}
c=E(m)=(g^x,mh^x)=(g^x,m{g^{xy}})=(c_1,c_2),
\end{equation}
\item \textit{Decryption Algorithm:} To decrypt the ciphertext $c$, first, $s={c_1}^{y}$ is computed where $y$ is the secret key. Then, decryption algorithm works as follows:
\begin{equation}
c_2\cdot s^{-1}=m{g^{xy}} \cdot g^{-xy}=m.
\end{equation}
\item \textit{Homomorphic Property:}
\begin{equation}
\begin{split}
E(m_1)*E(m_2)  =(g^{x_1},m_1{h^{x_1}})*(g^{x_2},m_2{h^{x_2}}) =(g^{{x_1}+{x_2}},{{m_1}*{m_2}}h^{{x_1}+{x_2}}) =E(m_1*m_2).
\end{split}
\end{equation}

\end{itemize}
\par As seen from this derivation, the \revision{El-Gamal} cryptosystem is multiplicatively homomorphic. It does not support addition operation over ciphertexts.
\subsubsection{Benaloh}
Benaloh proposed an extension of the GM Cryptosystem by improving it to encrypt the message as a block instead of bit by bit \cite{benaloh1994dense}. Benaloh's proposal was based on the higher residuosity problem. Higher residuosity problem ($x^n$)~\revision{\cite{Zheng_cryptographicapplications}} is the generalization of quadratic residuosity problems ($x^2$) that is used for the GM cryptosystem.
\begin{itemize}

\item \textit{KeyGen Algorithm:} Block size $r$ and large primes $p$ and $q$ are chosen such that r divides $p-1$ and r is relatively prime to $(p-1)/r$ and $q-1$ (i.e., $gcd(r,(p-1)/r)=1$ and $gcd(r,(q-1))=1$). Then, $n=pq$ and $ \phi=(p-1)(q-1)$ are computed. Lastly, $y \in {\mathbb{Z}_n}^* $ is chosen such that $y^\phi \not\equiv 1 \mod n$, where ${\mathbb{Z}_n}^*$ is the multiplicative subgroup of integers modulo $n$ which includes all the numbers smaller than $r$ and relatively prime to $r$. Finally, $(y,n)$ is published as the public key, and $(p,q)$ is kept as the secret key.
\item \textit{Encryption Algorithm:}
For the message $m \in Z_r$, where $Z_r=\{0,1,...,r-1\}$, choose a random $u$ such that $u \in {Z_n}^*$. Then, to encrypt the message $m$:
\begin{equation}
c=E(m)=y^mu^r \pmod n ,
\end{equation}
where the public key is the modulus $n$ and base $y$ with the block size of $r$.
\item \textit{Decryption Algorithm:} The message $m$ is recovered by an exhaustive search for $i \in {\mathbb{Z}_r}$ such that
\begin{equation}
(y^{-i}c)^{{\phi}/r} \equiv 1 ,
\end{equation}
where the message $m$ is returned as the value of $i$, i.e., $m=i$.
\item \textit{Homomorphic Property:}
\begin{equation}
\begin{split}
E(m_1)*E(m_2) & =({y^{m_1}}{u_1}^r \pmod n)*({y^{m_2}}{u_2}^r \pmod n)\\
& =y^{m_1+m_2}({u_1}*{u_2})^r \pmod n=E({m_1+m_2} \pmod n) .
\end{split}
\end{equation}
\end{itemize}

\par Homomorphic property of Benaloh shows that any multiplication operation on encrypted data corresponds to the addition on plaintext. As the encryption of the addition of the messages can directly be calculated from encrypted messages $E(m_1)$ and $E(m_2)$, the Benaloh cryptosystem is additively homomorphic.

\subsubsection{Paillier}
In 1999, Paillier~\cite{paillier1999public} introduced another novel probabilistic encryption scheme based on \textit{composite residuosity problem}~\revision{\cite{Jager2012}}. \textit{Composite residuosity problem} is very similar to quadratic and higher residuosity problems that are used in GM and Benaloh cryptosystems. It questions whether there exists an integer $x$ such that $x^n \equiv a \pmod {n^2}$ for a given integer $a$.
\begin{itemize}

\item \textit{KeyGen Algorithm:} For large primes $p$ and $q$ such that $gcd(pq,(p-1)(q-1))=1$, compute $n=pq$ and $\lambda=lcm(p-1,q-1)$. Then, select a random integer $g \in {{\mathbb{Z}^*}_{n^2}} $ by checking whether $gcd(n, L(g^{\lambda \mod {n^2}}))=1$, where the function $L$ is defined as $L(u)=(u-1)/n$ for every $u$ from the subgroup ${Z^*_{n^2}}$ which is a multiplicative subgroup of integers modulo $n^2$ instead of $n$ like in the Benaloh cryptosystem. Finally, the public key is $(n,g)$ and the secret key is $(p,q)$ pair.
\item \textit{Encryption Algorithm:}
\par For each message $m$, the number $r$ is randomly chosen and the encryption works as follows:
\begin{equation}
c=E(m)={g^m}{r^n} \pmod {n^2} ,
\end{equation}
\item \textit{Decryption Algorithm:}
For a proper ciphertext  $c<n^2$, the decryption is done by:
\begin{equation}
D(c)= \frac{L(c^ \lambda \pmod {n^2})}{L(g^ \lambda \pmod {n^2})} \mod n=m ,
\end{equation}
 where private key pair is $(p,q)$.
\item \textit{Homomorphic Property:}
\begin{equation}
\begin{split}
E(m_1)*E(m_2) & =(g^{m_1}{r_1}^n \pmod {n^2})\quad *(g^{m_2}{r_2}^n \pmod {n^2})\\ & =g^{m_1+m_2}(r_1*r_2)^n \pmod {n^2} =E(m_1+m_2) .
\end{split}
\end{equation}
\end{itemize}

\par This derivation shows that Pailliler's encryption scheme is homomorphic over addition. In addition to homomorphism over the addition operation, Pailliler's encryption scheme has some additional homomorphic properties, which allow extra basic operations on plaintexts $m_1,m_2 \in {Z^*_{n^2}}$ by using the encrypted plaintexts $E(m_1)$, $E(m_2)$ and public key pair $(n,g)$:
\begin{equation}\label{eq:1}
E(m_1)*E(m_2)\pmod {n^2}=E(m_1+m_2 \pmod n) ,
\end{equation}
\begin{equation}\label{eq:2}
E(m_1)*g^{m_2} \pmod {n^2}=E(m_1+m_2 \pmod n) ,
\end{equation}
\revision{
\begin{equation}\label{eq:3}
E(m_1)^{m_2} \pmod {n^2}=E({m_1}{m_2} \pmod n) .
\end{equation}
}
%\begin{equation}\label{eq:3}
%E(m)^k \pmod {n^2}=E(km \pmod n) .
%\end{equation}
\par These additional homomorphic properties describe different cross-relation between \revision{various} operations on the encrypted data and the plaintexts. In other words, Equations~\eqref{eq:1},~\eqref{eq:2}, and~\eqref{eq:3} show how \revision{the operations} computed on encrypted data affects the plaintexts.
\subsubsection{Others}
Moreover, Okamoto-Uchiyama (OU)~\cite{okamoto1998new} proposed a new PHE scheme to improve the computational performance by changing the set, where the encryptions of previous HE schemes work. The domain of the scheme is the same as the previous public key encryption schemes, ${Z^*_n}$, however, Okamoto-Uchiyama sets $n={p^2}q$ for large primes $p$ and $q$. Furthermore, Naccache-Stern (NS)~\cite{naccache1998new} presented another PHE scheme as a generalization of Benaloh cryptosystem to increase its computational efficiency. The proposed work changed only the decryption algorithm of the scheme. Likewise, Damgard-Jurik (DJ)~\cite{damgaard2001generalisation} introduced another PHE scheme as a generalization of Paillier. These three cryptosystems preserve the homomorphic property while improving the original homomorphic schemes.

\par Similarly, Kawachi (KTX) et al.~\cite{kawachi2007multi} suggested an additively homomorphic encryption scheme over a large cyclic group, which is based on the hardness of underlying lattice problems. They named the homomorphic property of their proposed scheme as \textit{pseudohomomorphic}. Pseudohomomorphism is an algebraic property and still allows homomorphic operations on ciphertext, however, the decryption of the homomorphically operated ciphertext works with a small decryption error. Finally, Galbraith~\cite{galbraith2002elliptic} introduced a more natural generalization of Paillier's cryptosystem applying it on elliptic curves while still preserving the homomorphic property of the Paillier's cryptosystem. Homomorphic properties of well-known PHE schemes are briefly summarized in Table~\ref{PHE_table}.

\begin{table}[t]
\centering
\caption{Homomorphic properties of well-known PHE schemes \label{PHE_table}}
\resizebox{10.5cm}{!}{
\begin{tabular}{|>{\small}l|>{\small}C{2cm}|>{\small}C{2cm}|}
\hline  & \multicolumn{2}{|c|}{{\small \textbf{Homomorphic Operation}}}\\
\hline \textbf{Scheme} & \textbf{Add} &  \textbf{Mult}\\ \hline \hline
RSA~\cite{rivest1978method} &  & \cmark  \\ 
GM~\cite{goldwasser1982probabilistic} & \cmark &   \\ 
\revision{El-Gamal}~\cite{elgamal1985public}\tablefootnote{\revision{The method to convert \revision{El-Gamal} into an additively homomorphic encryption scheme is shown in~\cite{cramer1997secure}. However, it is still PHE as it still supports only addition operation, not both at the same time.}} &  & \cmark  \\ 
Benaloh~\cite{benaloh1994dense} & \cmark &   \\ 
NS~\cite{naccache1998new} & \cmark &  \\ 
OU~\cite{okamoto1998new} & \cmark &    \\ 
Paillier~\cite{paillier1999public} &  \cmark &   \\ 
DJ~\cite{damgaard2001generalisation} & \cmark &   \\ 
KTX~\cite{kawachi2007multi} & \cmark &   \\ 
Galbraith~\cite{galbraith2002elliptic} & \cmark &   \\ 
\hline
\end{tabular}}
\end{table}

\subsection{Somewhat Homomorphic Encryption Schemes}
%In this section, we explain the SWHE schemes in detail.
There are useful SWHE examples~\cite{yao1982protocols,814630,boneh2005evaluating,ishai2007evaluating} in the literature before 2009. After the first plausible FHE published in 2009~\cite{gentry2009fully}, some SWHE versions of FHE schemes were also proposed because of the performance issues associated with FHE schemes. We cover \revision{these SWHE} schemes under the FHE section. In this section, we primarily focus on major SWHE schemes, which were used as a stepping stone to the first plausible FHE scheme.
\subsubsection{BGN}
Before 2005, all proposed cryptosystems' homomorphism properties were restricted to only either addition or multiplication operation i.e., SWHE schemes. One of the most significant steps toward an FHE scheme was introduced by Boneh-Goh-Nissim (BGN) in~\cite{boneh2005evaluating}. BGN evaluates 2-DNF\footnote{Disjunctive Normal Form with at most 2 literals in each clause.} formulas on ciphertext and it supports an arbitrary number of additions and one multiplication by keeping the ciphertext size constant. The hardness of the scheme is based on the \textit{subgroup decision problem}~\cite{gjosteen2004subgroup}. Subgroup decision problem simply decides whether an element is a member of a subgroup $G_p$ of group $G$ of composite order $n=pq$, where $p$ and $q$ are distinct primes.
\begin{itemize}
\item \textit{KeyGen Algorithm:} The public key is released as $(n,G,G_1,e,g,h)$. In the public key, $e$ is a bilinear map such that $e:G \times G \rightarrow G_1$, where $G,G_1$ are groups of order $n=q_1q_2$. $g$ and $u$ are the generators of $G$ and set $h=u^{q_2}$  and $h$ is the generator of $G$ with order $q_1$, which is kept hidden as the secret key.
\item \textit{Encryption Algorithm:} 
To encrypt a message $m$, a random number $r$ from the set $\{0,1,...,n-1\}$ is picked and encrypted using the precomputed $g$ and $h$ as follows:
\begin{equation}
c=E(m)={g^m}{h^r} \mod n
\end{equation}
\item \textit{Decryption Algorithm:} To decrypt the ciphertext $c$, one firstly computes $c'=c^{q_1}=({g^m}{h^r})^{q_1}=(g^{q_1})^m$ (Note that $h^{q_1} \equiv 1 \mod n$) and $g'=g^{q_1}$ using the secret key $q_1$ and decryption is completed as follows:
\begin{equation}
m=D(c)=\log_{g'} {c'}
\end{equation}
\par  In order to decrypt efficiently, the message space should be kept small because of the fact that discrete logarithm can not be computed quickly.

\item \textit{Homomorphism over Addition:} Homomorphic addition of plaintexts $m_1$ and $m_2$ using ciphertexts $E(m_1)=c_1$ and $E(m_2)=c_2$ are performed as follows:
\begin{equation}
\begin{split}
c ={c_1}{c_2}{h^r} =({g^{m_1}}{h^{r_1}})({g^{m_2}}{h^{r_2}}){h^r} =g^{m_1+m_2}{h^{r'}} ,
\end{split}
\end{equation}
where $r=r_1+r_2+r$ and it can be seen that $m_1+m_2$ can be easily recovered from the resulting ciphertext $c$.
\item \textit{Homomorphism over Multiplication:} To perform homomorphic multiplication, use $g_1$ with order $n$ and $h_1$ with order $q_1$ and set $g_1=e(g,g)$, $h_1=e(g,h)$, and $h=g^{\alpha {q_2}}$. Then, the homomorphic multiplication of messages $m_1$ and $m_2$ using the ciphertexts $c_1=E(m_1)$ and $c_2=E(m_2)$ are computed as follows:
\begin{equation}
\begin{split}
c & =e(c_1,c_2){{h_1}^r}=e({g^{m_1}}{h^{r_1}} ,{g^{m_2}}{h^{r_2}} ){h_1}^r\\
& ={{g_1}^{m_1m_2}}{{h_1}^{m_1r_2+r_2m_1+\alpha{q_2r_1r_2}+r}} ={{g_1}^{m_1m_2}}{{h_1}^{r'}} 
\end{split}
\end{equation}
\end{itemize}

\par It is seen that $r'$ is uniformly distributed like $r$ and so $m_1m_2$ can be correctly recovered from resulting ciphertext $c$. However, $c$ is now in the group $G_1$ instead of $G$. Therefore, another homomorphic multiplication operation is not allowed in $G_1$ because there is no pairing from the set $G_1$. However, resulting ciphertext in $G_1$ still allows an unlimited number of homomorphic additions. Moreover, Boneh et al. also showed the evaluation of 2-DNF formulas using the basic 2-DNF protocol. Their protocol gives a quadratic improvement in terms of the protocol complexity over Yao's well-known garbled circuit protocol in~\cite{yao1982protocols}.
\subsubsection{Others}

\revision{In the literature of HE schemes, one of the first SWHE schemes is Polly Cracker scheme~\cite{fellows1994combinatorial}. It allows both multiplication and addition operation over the ciphertexts. However, the size of the ciphertext grows exponentially with the homomorphic operation, especially multiplication operation is extremely expensive. Later more efficient variants~\cite{levy2004polly,van2006polly} are proposed, but almost all of them are later shown vulnerable to attacks~\cite{steinwandt2010ciphertext,levy2009survey}. Therefore, they are either insecure or impractical
%or lose the homomorphic property
~\cite{le2003polly}. Recently, \cite{albrecht2011polly} introduced a Polly Cracker with Noise cryptosystem, where the homomorphic addition operations do not increase the ciphertext size while the multiplications square it.}

\begin{table}[t]
\centering
\caption{Comparison of some well-known SWHE schemes before Gentry's work \label{SWHE}}
\resizebox{11cm}{!}{
\begin{tabular}{|>{\normalsize }l|>{\large }L|>{\large }c|>{\large }L|}
\hline
 & \textbf{Evaluation Size} & \textbf{Evaluation Circuit} & \textbf{Ciphertext Size} \\ \hline \hline
Yao~\cite{yao1982protocols} & arbitrary & garbled circuit & grows at least linearly \\ \hline
SYY~\cite{814630} & polly-many AND \& one OR/NOT  & $NC^1$ circuit & grows exponentially \\ \hline
BGN~\cite{boneh2005evaluating} & unlimited add \& 1 mult & 2-DNF formulas & constant \\ \hline
IP~\cite{ishai2007evaluating}& arbitrary & branching programs & doesn't depend on the size of function \\ \hline
\end{tabular}
}
%\vspace{-2mm}
\end{table}

Another idea of evaluating operations on encrypted data is realized over different sets. Sander, Young, and Yung (SYY) described first  SWHE scheme over a semi-group, $NC^1$,\footnote{NC stands for "Nick's Class" for the honor of Nick Pippenger}~\cite{814630}, which requires less properties than a group. $NC^1$ is a complexity class which includes the circuits with poly-logarithmic depth and polynomial size. The proposed scheme supported polynomially many ANDing of ciphertexts with one OR/NOT gate. However, the ciphertext size increased by a constant multiplication with each OR/NOT gate evaluation. This increase limits the evaluation of circuit depth. Yuval Ishai and Anat Paskin (IP) expanded the set to branching programs (aka Binary Decision Diagrams), which are the directed acyclic graphs where every node have two outgoing edges with labeled binary 0 and 1~\cite{ishai2007evaluating}. In other words, they proposed a public key encryption scheme by evaluating the branching programs on the encrypted data. Moreover, Melchor et al.~\cite{melchor2010additively} proposed a generic construction method to obtain a chained encryption scheme allowing the homomorphic evaluation of constant depth circuit over ciphertext. The chained encryption scheme is obtained from well-known encryption schemes with some homomorphic properties. For example, they showed how to obtain a combination of BGN~\cite{boneh2005evaluating} and Kawachi et al.~\cite{kawachi2007multi}. As mentioned before, BGN allows an  arbitrary number of additions and one multiplication while Kawachi's scheme is only additively homomorphic. Hence, the resulting combined scheme allows arbitrary additions and two multiplications. They also showed \revision{how this procedure is applied to the scheme in}~\cite{melchor2008lattice} allowing a  predefined number of homomorphic additions, to obtain a scheme which allows an arbitrary number of multiplications as well. However, in multiplication, ciphertext size grows exponentially while it is constant in a homomorphic addition. The summary of some well-known SWHE schemes is given in Table \ref{SWHE}. As shown in Table \ref{SWHE}, while in Yao, SYY, and IP cryptosystems, the size of the ciphertext grows with each homomorphic operation, in BGN it stays constant. This property of BGN is a significant improvement to obtain an FHE scheme. Accordingly, Gentry, Halevi, and Vaikuntanathan later simplified the BGN cryptosystem~\cite{gentry2010simple}. In their version, the underlying security assumption is changed to hardness of the LWE problem. The BGN cryptosystem chooses input from a small set to decrypt correctly. In contrast, a recent scheme introduced in~~\cite{gentry2010simple} have much larger message space. Moreover, some of the attempts to obtain an FHE scheme based on SWHE schemes are reported as broken. For instance, vulnerabilities for~\cite{mullen1994finite,i1996new,grigoriev2006homomorphic,domingo2002provably} were reported in~\cite{steinwandt2002cryptanalysis,choi2007cryptanalysis,wagner2003cryptanalysis,cheon2006known}, respectively.

\subsection{Fully Homomorphic Encryption Schemes}
\par An encryption scheme is called Fully Homomorphic Encryption (FHE) scheme if it allows an unlimited number of evaluation operations on the encrypted data and resulting output is within the ciphertext space. After almost 30 years from the introduction of privacy homomorphism concept~\cite{rivest1978data}, Gentry presented the first feasible proposal in his seminal PhD thesis to a long term open problem, which is obtaining an FHE scheme \cite{gentry2009fully}. Gentry's proposed scheme gives not only an FHE scheme, but also a general framework to obtain an FHE scheme. Hence, a lot of researchers have attempted to design a secure and practical FHE scheme after Gentry's work.

\begin{figure}[t]
\centering
\begin{tikzpicture}[sibling distance=10em,font=\fontsize{6}{8}\selectfont,
  a/.style = {shape=rectangle, rounded corners, draw, align=center, top color=white, bottom color=blue!0, minimum width={width("FHE")+25pt}, minimum height = {height("FHE")+15pt}},   b/.style = {shape=rectangle, rounded corners, draw, align=center, top color=white, bottom color=blue!0}]
  \node[a] {FHE}
    child[b] { node[b] {Ideal Lattice-based\\ \cite{gentry2009fully} }}
    child[b] { node[b] {Over Integers\\ \cite{van2010fully}}}
    child[b] { node[b] {(R)LWE-based\\ \cite{brakerski2011fully}}}
    child[b] { node[b] {NTRU-like\\ \cite{lopez2012fly}} };
\end{tikzpicture}
\caption{Main FHE families after Gentry's breakthrough}
\label{fhegroups}
\end{figure}
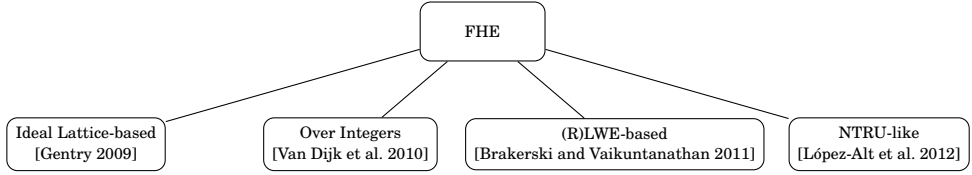 

\par Although Gentry's proposed ideal lattice-based FHE scheme~\cite{gentry2009fully} is very promising, it also had a lot of bottlenecks such as its computational cost in terms of applicability in real life and some of its  advanced mathematical concepts make it
%difficult for average users to understand. 
\revision{complex and hard to implement.}
Therefore, many new schemes and optimization have followed his work in order to address aforementioned bottlenecks. The security of new approaches to obtain a new FHE scheme is mostly based on \revision{the hard} problems on lattices. 
\par A lattice is the linear combinations of independent vectors (basis vectors), $b_1,b_2,...,b_n$. A lattice $L$ is formulated as follows:
 \begin{equation}
 L=\displaystyle\sum_{i=1}^{n} \vec{b_i} * {v_i} \quad, v_i \in \mathbb{Z},
 \end{equation}
where each vectors $b_1,b_2,...,b_i$ is called a basis of the lattice $L$. The basis of a lattice is not unique. There are infinitely many bases for a given lattice. A basis is called "good" if the basis vectors are almost orthogonal and, otherwise it is called "bad" basis of the lattice\revision{~\cite{micciancio2009lattice}}. Roughly, while good bases are typically long, bad bases are relatively shorter. Indeed, the lattice theory is firstly presented by Minkowski \cite{minkowski1968geometrie}. Then as a seminal work, Ajtai mentioned a class of random worst-case lattice problem in \cite{ajtai1996generating}. Two well-known modern problems suggested in~\cite{ajtai1996generating} for lattice-based cryptosystems are Closest Vector Problem (CVP) and Shortest Vector Problem (SVP) \cite{peikert2015decade}. A year after, Goldreich, Goldwasser, and Halevi (GGH) \cite{goldreich1997public} proposed an important type of PKE scheme, whose hardness is based on \textit{the lattice reduction problems} \cite{peikert2015decade}. Lattice reduction tries to find a good basis, which is relatively short and orthogonal, for a given lattice. In GGH cryptosystem, the public key and the secret key is chosen from "bad" and "good" basis of the lattice, respectively. The idea behind this choice is that CVP and SVP problems can easily be solved in polynomial time for the lattices with the known good bases. However, best known algorithms (for example LLL in \cite{lenstra1982factoring}) solve these problems in exponential time without knowing the good bases of the lattice. Hence, recovering the message from a given ciphertext is equal to \revision{solving} the CVP and SVP problems.
%in a practical time
%Then, 
\revision{In GGH cryptosystem, the message is embedded to the noise
%. Finally, the noise is added to a lattice point 
to obtain the ciphertext. In order to recover the message from ciphertext, the secret key (good basis) is used to find the closest lattice point.}

Before Gentry's work, in \cite{regev2006lattice}, cryptographers' attention is drawn to lattice-based cryptology and especially its great promising properties for post-quantum cryptology. Its promising properties are listed as its security proofs, efficient implementations, and simplicity. Moreover, another lattice-related problem, which gains popularity in last few years, especially after being used as a base to built an FHE scheme is LWE \cite{zhang2014revisiting}.
%In addition, 
One of the most significant works for lattice-based cryptosystems was studied in \cite{hoffstein1998ntru}\revision{, which} presented a new PKE scheme and \revision{whose} security is based on SVP on the lattice. In the SVP problem, given a basis of a lattice, the goal is to find the shortest nonzero vector in the lattice.

\par After Gentry's work, the lattices have become more popular among cryptography researchers. %The follow-up works can be categorized under four main FHE families in terms of the hardness of the problems, on which the FHE scheme is based: 
First, some works like~\cite{smart2010fully}
focused on just improving Gentry's ideal lattice-based FHE scheme in~\cite{gentry2009fully}. Then,  an FHE scheme over integers based on the Approximate-GCD problems is introduced \cite{van2010fully}. The main motivation behind the scheme is the conceptual simplicity. Afterwards, another FHE scheme whose hardness based on Ring Learning with Error (RLWE) problems is suggested \cite{brakerski2011fully}. The proposed scheme promises some efficiency features. Lastly, an NTRU-like FHE is presented  for its promising efficiency and  standardization properties \cite{lopez2012fly}. NTRUEncrypt is an old and strongly standardized lattice-based encryption scheme whose homomorphic properties are realized recently. So, these and similar attempts can be categorized into under four main FHE families as shown in Figure \ref{fhegroups}: (1) Ideal lattice-based~\cite{gentry2009fully}, (2) Over integers~\cite{van2010fully}, (3) (R)LWE-based~\cite{brakerski2011fully}, and (4) NTRU-like~\cite{lopez2012fly}. In the following sections, we will articulate these four main FHE families in greater detail. And, we will also explore other follow-up works after these. % Is this modification ok? COMMENT 
%following works of these four main schemes will be explained.

\subsubsection{ Ideal Lattice-based FHE schemes} 
Gentry's first FHE scheme in his PhD thesis~\cite{gentry2009fully} is a GGH-type of encryption scheme\revision{, where GGH is }proposed originally by Goldreich et al. \cite{goldreich1997public}. However, Gentry encrypted the message by embedding noise using double layer instead of one layer idea in GGH cryptosystem. Indeed, Gentry started his breakthrough work from SWHE scheme based on ideal lattices. 

\par As mentioned earlier, an SWHE scheme can evaluate the ciphertext homomorphically for only a limited number of operations. After a certain threshold, the decryption function fails to recover the message from the ciphertext correctly. The amount of noise in the ciphertext must be decreased to transform the noisy ciphertext into a proper ciphertext. Gentry used genius blueprint methods called  \textit{squashing}  and \textit{bootstrapping} to obtain a ciphertext which allows a number of homomorphic operations to be performed on it. This processes can be repeated again and again. In other words, one can evaluate unlimited operations on the ciphertexts which make the scheme fully homomorphic.

\par As an initial construction, Gentry used ideals and rings without lattices to design the homomorphic encryption scheme, where an ideal is a property preserving subset of the rings such as even numbers. Then, each ideal used in his scheme was represented by the lattices. For example, an ideal $I$ in $\mathbb{Z}[x]/(f(x))$ with $f(x)$ of degree $n$ in an ideal lattice can easily be represented by a column of lattice with basis $B_I$ of length $n$. Since the bases $B_I$ will produce an $n \times n$ matrix. Gentry's SWHE scheme using ideals and rings is described below:
\begin{itemize}

\item \textit{KeyGen Algorithm:} For the given ring $R$ and the basis $B_I$ of ideal $I$, $IdealGen(R,B_I)$ algorithm generates the pair of $(B_{J}^{sk},B_{J}^{pk})$, where $IdealGen()$ is an algorithm outputting the relatively prime public and the secret key bases of the ideal lattice with basis $B_I$ such that $I+J=R$. A $Samp()$ algorithm is also used in key generation to sample from the given coset of the ideal, where a coset is obtained by shifting an ideal by a certain amount. Finally, the public key consists of $(R,B_I,B_{J}^{pk},Samp())$ and the secret key only includes $B_{J}^{sk}$. 

\item \textit{Encryption Algorithm:}
\par For randomly chosen vectors $\vec{r}$ and $\vec{g}$, using the public key (basis) $B_{pk}$ chosen from one of the "bad" bases of the ideal lattice L, the message  $\vec{m} \in {\{0,1\}}^n $ is encrypted by:\\
\begin{equation}
\vec{c}=E(\vec{m})=\vec{m}+\vec{r} \cdot {B_I} + \vec{g} \cdot B_{J}^{pk},
\end{equation}\\
where $B_I$ is basis of the ideal lattice $L$. Here, $\vec{m}+\vec{r} \cdot {B_I}$ is called "noise" parameter.
\item \textit{Decryption Algorithm:}
\par By using the secret key (basis) $B_{J}^{sk}$, the ciphertext is decrypted as follows:
\begin{equation}
\vec{m}=\vec{c}-B_{J}^{sk} \cdot \lfloor (B_{J}^{sk})^{-1}\cdot \vec{c} \rceil\mod{B_I},
\end{equation}
where $\lfloor \cdot \rceil$ is the \textit{nearest integer function} which returns the nearest integers for the coefficients of the vector.

\item \textit{Homomorphism over Addition:} For the plaintext vectors  $\vec{m_1},\vec{m_2} \in \{0,1\}^n $, additive and multiplicative homomorphisms can be verified easily as follows:
\begin{equation}
 \vec{c_1}+\vec{c_2}=E(\vec{m_1})+E(\vec{m_2})=\vec{m_1}+\vec{m_2}+(\vec{r_1}+\vec{r_2}) \cdot {B_I}+ (\vec{g_1}+\vec{g_2}) \cdot B_{J}^{pk}
\end{equation}

\par It is clear that $\vec{c_1}+\vec{c_2}$ still preserves the format and is within the ciphertext space. And, to decrypt the sum of the ciphertext, one computes $(\vec{c_1}+\vec{c_2}) \mod B_{J}^{pk}$ which is equal to $\vec{m_1}+\vec{m_2}+(\vec{r_1}+\vec{r_2}) \cdot {B_I}$ for the ciphertexts whose noise amount is smaller than $B_{J}^{pk}/2$. Then the decryption algorithm works properly and recovers the sum of the message $m_1+m_2$ correctly by taking the modulo $B_I$ of the noise.

\item \textit{Homomorphism over Multiplication:} Similarly for the multiplication, after setting $ \vec{e}=\vec{m}+\vec{r} \cdot B_I$, the homomorphic property can be expressed as follows:
\begin{equation}
\vec{c_1} \times \vec{c_2}=E(\vec{m_1}) \times E(\vec{m_2})=\vec{e_1} \times \vec{e_2}+(\vec{e_1} \times \vec{g_2}+\vec{e_2} \times \vec{g_1}+\vec{g_1} \times \vec{g_2}) \cdot B_{J}^{pk}
\end{equation}
where $\vec{e_1} \times \vec{e_2}=\vec{m_1} \times \vec{m_2}+(\vec{m_1} \times \vec{r_2}+\vec{m_2} \times \vec{r_1}+\vec{r_1} \times \vec{r_2}) \cdot B_I$. It can be easily verified that the multiplication operation on ciphertexts yields the output still within the ciphertext space. It is said that if the noise $|\vec{e_1}\times\vec{e_1}|$ is enough small enough the multiplication of plaintexts $\vec{m_1} \times \vec{m_2}$  can be correctly recovered from the multiplication of ciphertexts $\vec{c_1} \times \vec{c_2}$.
\end{itemize}

\par \revision{To have a better understanding of the "noise" concept, let us consider the encryption scheme over integers\footnote{\revision{Further details about FHE over integers will be explained in Section~\ref{sec:fhe_over_integers}.}}. The encryption of the bit $b$ is the ciphertext $c=b+2r+kp$, where the key $p>2N$ is an odd integer and $r$ is a random number from the range $(-n/2,n/2)$ and $k$ is an integer. The decryption works as follows: $b \leftarrow (c \mod p) \mod 2$, where $(c \mod p)$ is called as \textit{noise parameter}. If the noise parameter exceeds $|p/2|$, the decryption fails since $(c \mod p)$ is not equal to $b+2r$ anymore.}
%\par As seen above, homomorphic operations can be applied to only ciphertexts with small amount of noises. \revision{Otherwise decryption would fail. In other words,} if the noise parameter is very close to a lattice point, further addition and multiplication operations are still allowed. However, after a threshold point, it is not possible to decrypt the ciphertext properly.
And, the noise parameter grows linearly with each addition and exponentially with each multiplication operation. \revision{If the noise parameter is very close to a lattice point (i.e., $(c \mod p) << |p/2|$), further addition and multiplication operations are still allowed.} This is why \revision{Gentry's ideal lattice based} scheme is called Somewhat Homomorphic "for now" allowing only limited number of operations. Since the noise grows much faster with the multiplication operations, the number of multiplication operations before exceeding the threshold is more limited. In order to make the scheme fully homomorphic, the bootstrapping technique was introduced by Gentry. However, the bootstrapping process can be applied to the bootstrappable ciphertexts, which are noisy and have small circuit depth. The depth of the circuit is related to the maximum number of operations. Hence, first the circuit depth is reduced with \textit{squashing} to the degree that the decryption can handle properly. \\

 \textbf{Squashing:} Gentry's bootstrapping technique is allowed only for the decryption algorithms with small depth. Therefore, he used some "tweaks" to reduce the decryption algorithm's complexity. This method is called \textit{squashing} and works as follows:
 
\par First, choose a set of vectors, whose sum equals to the multiplicative inverse of the secret key ($(B_{J}^{sk})^{-1}$). If the ciphertext is multiplied by the elements of this set, the polynomial degree of the circuit is reduced to the level that the scheme can handle. The ciphertext is now "bootstrappable". Nonetheless, the hardness of the recovering the secret key is now based on the assumption of Sparse Subset Sum Problem (SSSP)~\cite{hoffstein2008introduction}. This basically adds another assumption to the provable security of the scheme.\\

\textbf{Bootstrapping:} Bootstrapping is basically "recrypting" procedure to get a "fresh" ciphertext from the noisy ciphertext corresponding to the same plaintext. A scheme is called \textit{bootstrappable}  if it can evaluate its own decryption algorithm circuit~\cite{gentry2009fully}. First, the ciphertext is transformed into a bootstrappable ciphertext using squashing. Then, by applying bootstrapping procedure, one gets a "fresh" ciphertext. The bootstrapping works as follows: First, it is assumed that two different public and secret key pairs are generated, $(pk1,sk1)$ and $(pk2,sk2)$ and while the secret keys are kept by the client, the public keys are shared with the server. Then, the encryption of the secret key, $Enc_{pk1}(sk1)$, is also transmitted to the server, which already has $c=Enc_{pk1}(m)$. Since the above obtained SWHE scheme can evaluate its own decryption algorithm homomorphically, the noisy ciphertext is decrypted homomorphically using $Enc_{pk1}(sk1)$. Then, the result is encrypted using a different public key $pk2$, i.e., $Enc_{pk2}(Dec_{sk1}(c))=Enc_{pk2}(m)$. Since the scheme is assumed semantically secure, an adversary can not distinguish the encryption of the secret key from the encryption of $0$. The last ciphertext can be decrypted using $sk2$, which is kept secret by the client, i.e., $Dec_{sk2}(Enc_{pk2}(m))=m$. In brief, first the homomorphic decryption of the noisy ciphertext removes the noise, and then the new homomorphic encryption introduces new small noise to the ciphertext. Now, the ciphertext is like just encrypted. Further homomorphic operations can be computed on this "fresh" ciphertext until reaching again to a threshold point. Note that Gentry's bootstrapping method increases the computational cost noticeably and becomes a major drawback for the practicality of FHE. In a nutshell, starting from constructing a SWHE scheme and then squashing method to reduce the circuit depth of decryption algorithm and the bootstrapping to obtain fresh ciphertext completes the creation of an FHE scheme. Hence, one can apply bootstrapping repetitively to compute an unlimited number of operations on the ciphertexts to successfully have an FHE scheme.

\par After Gentry's original scheme, some of the follow-up works tried to generally improve Gentry's original work. In \cite{gentry2009fully}, Gentry's key generation algorithm is used for a particular purpose only and the generation of an ideal lattice with a "good" basis is left without a solution. Gentry introduced a new $KeyGen$ algorithm in \cite{gentry2010toward} and improved the security of the hardness assumption of SSSP by presenting a quantum worst case/average case reduction. However, a more aggressive analysis of the security of SSSP was completed by Stehle and Steinfeld \cite{stehle2010faster}. They also suggested a new probabilistic decryption algorithm with lower multiplicative degree, which is square root of previous decryption circuit degree. Moreover, a new FHE scheme, which was a variant of Gentry's scheme was introduced in \cite{smart2010fully}. The scheme uses smaller ciphertext and key sizes than Gentry's scheme without sacrificing the security. Some later works \cite{gentry2011implementing,scholl2011improved,ogura2010improvement} focused on the optimizations in the key generation algorithm in order to implement the FHE efficiently. Moreover, Miku\v{s} proposed a new SWHE scheme with bigger plaintext space to improve the number of homomorphic operations with a slight increase in complexity of the key generation algorithm \cite{mikuvs2012experiments}.\\

\subsubsection{FHE schemes Over Integers} \label{sec:fhe_over_integers}
\par In 2010, one year after Gentry's original scheme, another SWHE scheme is presented in \cite{van2010fully} which suggests Gentry’s ingenious bootstrapping method in order to obtain an FHE scheme. The proposed scheme is over integers and the hardness of the scheme is based on the \textit{Approximate-Greatest Common Divisor} (AGCD) problems \cite{galbraithalgorithms}. AGCD problems try to recover $p$ from the given set of $x_i=p{q_i}+r_i$. The primary motivation behind the scheme is its conceptual simplicity. A symmetric version of the scheme is probably one of the simplest schemes. The proposed symmetric SWHE scheme is described as follows: 
\begin{itemize}
\item \textit{KeyGen Algorithm:} For the given security parameter $\lambda$, a random odd integer $p$ of bit length $\eta$ is generated.
\item \textit{Encryption Algorithm:} For a random large prime numbers $p$ and $q$, choose a small number $r<<p$. Then, the message $m \in \{0,1\} $ is encrypted by:
\begin{equation}
c=E(m)=m+2r+pq,
\end{equation}
where $p$ is kept hidden as private key and $c$ is the ciphertext.
\item \textit{Decryption Algorithm:} The ciphertext can be decrypted as follows:
\begin{equation}
m=D(c)=(c \mod p)\mod 2.
\end{equation}
Decryption works properly only if $m+2r<p/2$. This actually restricts the depth of the homomorphic operations performed on the ciphertext. Then, Dijk et al. used Gentry's  squashing and bootstrapping techniques to make the scheme fully homomorphic. The homomorphic properties of the scheme can be shown easily as follows:
\item \textit{Homomorphism over addition:}
\begin{equation}
E(m_1)+E(m_2)=m_1+2r_1+pq_1+m_2+2r_2+pq_2=(m_1+m_2)+2(r_1+r_2)+(q_1+q_2)q.
\end{equation}

The output clearly falls within the ciphertext space and can be decrypted if the noise $|m_1+2r_1+m_2+2r_2|<p/2$, where p is the private key. Since $r_1,r_2<<p$, various number of additions can still be performed on ciphertext before noise exceeds $p/2 $.
\item \textit{Homomorphism over Multiplication:}
\begin{equation}
E(m_1)E(m_2)=(m_1+2r_1+pq_1)(m_2+2r_2+pq_2=m_1m_2+2(m_1r_2+m_2r_1+2r_1r_2)+kp.
\end{equation}

The output preserves the format of original ciphertexts and holds the homomorphic property. The encrypted data can be decrypted if the  noise is smaller than half of the private key, i.e., $|m_1m_2+2(m_1r_2+m_2r_1+2r_1r_2)|<p/2$. The noise grows exponentially with the multiplication operation. This puts more restriction over homomorphic multiplication operation than addition.
\end{itemize}

\par In fact, the scheme presented so far \cite{van2010fully} was the symmetric version of the homomorphic encryption. Transforming the underlying symmetric HE scheme into an asymmetric HE scheme is also presented in \cite{van2010fully}. It is enough to compute many "encryptions of zero" $x_i=p{q_i}+2{r_i}$, where $p$ is private key. Then, many $x_i$s are shared as the public key. To encrypt the message with the public key, it is enough to add the message to a subset sum of $x_is$. Same decryption is used to decrypt the ciphertext. As there is no efficient algorithm to recover $p$ from the given $x_i$s in polynomial time, the scheme is considered as secure. The scheme is now basically a public key encryption scheme, since it uses different keys to encrypt and decrypt.

\par The FHE scheme proposed in \cite{van2010fully} is conceptually very simple. However, this simplicity comes at a cost in computations. So, the scheme is not very efficient. Hence, some early attempts directly tried to improve the efficiency. For example, some follow-up optimizations focused on reducing the size of public keys \cite{coron2011fully} ($O(\lambda^{10})\rightarrow O(\lambda^7) $), \cite{coron2012public} ($ O(\lambda^7)\rightarrow O(\lambda^5$), \cite{yang2012new} ($ O(\lambda^5)\rightarrow O(\lambda^3$). A more efficient public key generation \cite{ramaiah2012towards} and re-encryption~\cite{chen2014encryption} are other suggested works without reducing the security of the scheme. Later, an important variant, which is batch FHE over integers, was proposed \cite{cheon2013batch} (merged version of \cite{cryptoeprint:2013:036} and \cite{kim2013crt}). Batch FHE has the ability to pack multiple ciphertexts into a single ciphertext. Moreover, the proposed scheme provides two options for the hardness of the base problem: Decisional AGCD and Error-free AGCD. In \cite{cheon2013batch}, it is also shown how to achieve recryption operation in parallel $l$-slots. 

\par Some further approaches for FHE schemes over integers are also proposed: a new scale invariant FHE over integers \cite{coron2014scale}, a new scheme with integer plaintexts \cite{ramaiah2012efficient}, a new SWHE scheme for computing arithmetic operations on large integer numbers without converting them into bits \cite{pisa2012somewhat}, a new symmetric FHE without bootstrapping \cite{aggarwal2014fully}, and a new FHE for non-binary message spaces \cite{nuida2015batch}. All these schemes improved FHEs over integers in the way that their names imply.

\subsubsection{LWE-based FHE schemes}
Learning with Error (LWE) is considered as one of the hardest problems to solve in practical time for even post-quantum algorithms. First, it was introduced by Oded Regev as an extension of "learning from parity with error" problem \cite{regev2009lattices}. Regev reduced the hardness of worst-case lattice problems like SVP to LWE problems, which means that if one can find an algorithm that can solve LWE problem in an efficient time, the same algorithm will also solve the SVP problem in an efficient time.   Since then, it is one of the most attractive and promising topics for post-quantum cryptology with its relatively small ciphertext size. Lyubashevsky et al. suggested another significant improvement on the LWE problem which may lead to a new applications by introducing ring-LWE (RLWE) problem \cite{lyubashevsky2013ideal}. The RLWE problem is an algebraic variant of LWE, which is more efficient for practical applications with strong security proofs. They proved that the RLWE problems are reducible to worst-case problems on ideal lattices, which is hard for polynomial-time quantum algorithms.

\par In the LWE-based FHE schemes, an important step towards to a practical FHE scheme is made in \cite{brakerski2011fully}. Brakerski and Vaikuntanathan established a new SWHE scheme based on Ring-Learning with Error (RLWE) to take advantage of the efficiency feature of RLWE \cite{brakerski2011fully}. In other words, although both LWE and RLWE problems can be used as the hardness assumption of an FHE scheme, RLWE shows better performance. Then, the scheme uses Gentry's blueprint squashing and bootstrapping techniques to obtain an FHE scheme. They used polynomial-LWE (PLWE), which is simplified version of RLWE. PLWE is also reducible to worst-case problems such as SVP on ideal lattices. The schemes proposed after \cite{brakerski2011fully} is also called second generation FHE schemes. 

\par Below, for the sake of simplicity, as we did in the previous part, we first show symmetric version.
%of the scheme and then how to transform the scheme into a asymmetric HE scheme.\\

\textit{Notation:} A very common notation is that $\langle a,b \rangle$ is used to denote the inner product of vectors $a$ and $b$. Moreover, $d \xleftarrow{\$} \mathcal{D} $ denotes that $d$ is randomly assigned by an element from the distribution $ \mathcal{D} $ and $\mathbb{Z}[x]/(f(x))$ denotes the ring of all polynomials modulo $f(x)$. The ring of polynomials modulo $f(x)$ with coefficients in $\mathbb{Z}_q$ is denoted with $R_q \equiv \mathbb{Z}_q[x]/(f(x))$. Finally, $\chi$ denotes an error distribution over the ring $R_q$.

The symmetric version of the underlying scheme is given as follows:
\begin{itemize}
\item \textit{KeyGen Algorithm:} An element of the ring is chosen as a secret key from the error distribution, i.e., $s \xleftarrow{\$} \chi $. Then, the secret key vector is described as $\vec{s}=(1,s,s^2,...,s^D)$ for an integer $D$.
\item \textit{Encryption Algorithm:} After choosing a random vector $a \xleftarrow{\$} {R_q}^n$ and the noise $e \xleftarrow{\$} \chi $, the message $m$ is encrypted by:\\
\begin{equation}
\vec{c}=(c_0,c_1)=(as+te+m,-a)    
\end{equation}
where $\vec{c} \in R_q^2$.
\item \textit{Decryption Algorithm:} In order to decrypt the ciphertext to recover the message, it can be easily computed that:
\begin{equation}
m= \langle \vec{c},\vec{s} \rangle \pmod t .
\end{equation}
\par Decryption works properly if $\langle \vec{c},\vec{s} \rangle$ is smaller than $q/2$. Furthermore, in order to make the scheme asymmetric, it is sufficient to generate a random set of pairs $(a,as+te)$. Also, the homomorphic property of the scheme is very similar to those in \cite{gentry2009fully} and \cite{van2010fully}.
\item \textit{Homomorphism over Addition:}
\begin{equation}
E(m)+E(m')=(c_0+c'_0,c_1+c'_1)=((a+a')s+t(e+e')+(m+m'),-(a+a')),
\end{equation}
\par Similar to previous schemes, decryption works if the noise is small. And, it is clear that homomorpically added ciphertexts keep the format of the original ciphertexts and stay within the ciphertext space.
 \item \textit{Homomorphism over Multiplication:}
\begin{equation}
E(m)+E(m')=({c_0}{c'_0},{c_1}{c'_1})=(-a's^2+(c'_0a+c_0a')s+t(2ee'+em'+e'm)+mm').
\end{equation}
The output seems almost like a ciphertext, but it still can be decrypted correctly with the expense of a new cost by adding a new term to ciphertext.
\end{itemize}

\par Brakerski and Vaikuntanathan made their scheme fully homomorphic using Gentry's blueprint squashing and bootstrapping. They also showed their SWHE scheme is circular secure (aka Key-Dependent message (KDM) security) with respect to linear functions of the secret key, i.e., the encryption can successfully keep secure linear functions of its own secret key.

\par After the proposed BGN-type cryptosystem based on LWE, which is additively homomorphic and allowing only one multiplication operation in~\cite{gentry2010simple},  Brakerski and Vaikuntanathan proposed another SWHE scheme based on standard LWE problems using \textit{re-linearization} technique \cite{brakerski2014efficient}. Re-linearization makes the long ciphertexts, which are the output of the homomorphic evaluation, regular size. Another important contribution in this work is the dimension-modulus reduction, which does not require an SSSP assumption and squashing method used in Gentry's original framework. 

\par As discussed earlier, Gentry's bootstrapping method is a creative method to obtain an FHE scheme, however, it comes with a huge cost. A \textit{leveled}-FHE scheme without using the bootstrapping technique was introduced by \cite{brakerski2012leveled}. Levelled FHE can evaluate homomorphic operations for only a predetermined circuit depth level. Brakerski et al.~\cite{brakerski2012leveled} also showed that their scheme with bootstrapping still provides better performance than the one without bootstrapping and also suggested the batching as an optimization. To achieve batching, "modulus switching" technique is used iteratively to keep the noise size constant. Then, Brakerski removed the necessity of modulus switching in~\cite{brakerski2012fully}. In Brakerski's new scale invariant FHE scheme~\cite{brakerski2012fully}, contrary to the existing FHE schemes, the noise grows linearly with the evaluation of homomorphic operations instead of exponentially and the scheme is based on the hardness of \textit{GapSVP problem}~\cite{peikert2015decade}. GapSVP problem is roughly deciding the existence of a shorter vector than the vector with length $d$ for a given lattice basis $B$. The result returns simply yes or no. Then, Fan and Vercauteren optimized the Brakerski's scheme by changing the based assumption to RLWE problem~\cite{fan2012somewhat}. Some other modifications to~\cite{brakerski2012fully} focused on reducing the overhead of key switching  and faster evaluation of homomorphic operations~\cite{wu2012optimizations} and using re-linearization to improve efficiency~\cite{zhang2014efficient}.

\par Recently, by \cite{gentry2013homomorphic} a significant FHE scheme was introduced claiming three important properties: simpler, faster, and attribute-based FHE. The scheme is simpler and faster due to the "approximate eigenvector" method replacing the re-linearization technique. In this method, by keeping only some parameters small, the format of the ciphertext can be preserved under the evaluation of homomorphic operations. In the previous schemes which use the bootstrapping technique, the secret key (evaluation key) of the user is sent to the cloud to evaluate the ciphertext homomorphically for the bootstrapping. In contrast, \cite{gentry2013homomorphic} eliminates that need and leads to propose the first identity-based FHE scheme, which allows homomorphic evaluation by only a target identity having the public parameters. Then, Brakerski and Vaikuntanathan followed \cite{gentry2013homomorphic} to construct an FHE scheme secure under a polynomial LWE assumption \cite{brakerski2014lattice}. It is shown that the proposed scheme is as secure as any other lattice-based PKE scheme. Recently, Paindavoine and Vialla showed a way of minimizing the number of required bootstrapping based on the linear programming techniques that can be applied to \cite{gentry2013homomorphic} as well.
\par In addition to more recently proposed LWE-based FHE schemes in \cite{zhang2014efficient,chen2014fully,tanping1efficient,wang2015lwe}, some optimizations focused on better (faster) bootstrapping algorithms \cite{alperin2013practical,alperin2014faster}, speeding homomorphic operations \cite{gentry2012ring}, and a new extension to FHE for multi-identity and multi-key usage \cite{clearmulti}. More recently, a new efficient SWHE scheme based on the polynomial approximate common divisor problem is presented in \cite{cheon2016polynomial}. The presented scheme in \cite{cheon2016polynomial} can handle efficiently large message spaces.

\subsubsection{NTRU-like FHE schemes}
To obtain a practical and applicable FHE scheme, one of the crucial steps is taken by showing the construction of an FHE scheme from NTRUEncrypt, which is an old encryption scheme proposed by Hoffstein, Pipher, and Silvermanin in \cite{hoffstein1998ntru}. Specifically, how to obtain a multi-key FHE from the NTRUEncrypt (called NTRU) was shown by \cite{lopez2012fly}. NTRU encryption scheme is one of earliest attempts based on lattice problems. Compared with RSA and GGH cryptosystems, NTRU improves the efficiency significantly in both hardware and software implementations. However, there were security concerns for 15 years until the study done by \cite{stehle2011making}. They reduced the security of the scheme to standard worst-case problems over ideal lattices by modifying the key generation algorithm. Since the security of the scheme is improved, efficiency, easy implementation, and  standardization issues  attract researchers' interest again. At the same time, \cite{lopez2012fly} and \cite{test1} independently noticed the fully homomorphic properties of the NTRU encryption. López-Alt et al. used the NTRU encryption scheme to obtain a practical FHE \cite{lopez2012fly} with three differences. First, the set from which the noise is sampled is changed from a deterministic set to a distribution. Second, the modification introduced in~\cite{stehle2011making}, which makes the scheme more secure, is used and third, the parameters are chosen to allow fully homomorphism. Their proposed NTRU-like encryption scheme in \cite{lopez2012fly} is as follows:
\begin{itemize}

\item \textit{KeyGen Algorithm:} For chosen sampled polynomials $f'$ and $g$  from a distribution $\chi$ (specifically, a discrete
Gaussian distribution), it is set $f=2f'+1$ to get $f \equiv 1 \pmod 2$ and f is invertible. Then, the secret  key $sk=f \in R$ and public key $pk:=h=2g{f^{-1}} \in R_q$.
\item \textit{Encryption Algorithm:} For chosen samples $s$ and $e$ from the same distribution $\chi$, the message $m$ is encrypted by:
\begin{equation}
c=E(m)=hs+2e+m,
\end{equation}
where the ciphertext $c \in R_q$.
\item \textit{Decryption Algorithm:} The ciphertext can easily be decrypted as follows:
\begin{equation}
m=D(c)=fc \pmod 2,
\end{equation}
where $fc \in R_q$. The correctness of the scheme can be verified using $h=2g{f^{-1}}$ and $f \equiv 1 \pmod 2$. Moreover, the scheme proposed by L{\'o}pez-Alt et al. is a new type of FHE scheme, which is called multi-key FHE. Multi-key FHE has the ability to evaluate on ciphertexts which are encrypted with independent keys, i.e., each user can encrypt data with her own public key and a third party can still perform a homomorphic evaluation on these ciphertexts. The only interaction required between the users is to obtain a "joint secret key". The homomorphically evaluated ciphertext is decrypted by using the joint secret key, which is obtained by using all involved secret keys. The message $m_i$ is encrypted by using public key $h_i=2{g_i}{{f_i}^{-1}}$  with the formula, $c_i={h_i}{s_i}+2{e_i}+m_i$. The multikey homomorphism properties for two party computation is shown using joint secret key $f_1f_2$.
\item \textit{Multi-key Homomorphism over Addition:}
\begin{equation}
\begin{split}
{f_1}{f_2}(c_1 + c_2) = & 2 (f_1 f_2 e_1 + f_1f_2 e_2 + f_2 g_1s_1 + f_1 g_2 s_2 )+ f_1 f_2(m_1 + m_2 )\\
& = 2e_{add} + f_1f_2 (m_1 + m_2) 
\end{split}
\end{equation}
\item \textit{Multi-key Homomorphism over Multiplication:}
\begin{equation}
\begin{split}
f_1 f_2(c_1 c_2) = & 2(2g_1 g_2 s_1 s_2 + g_1 s_1 f_2(2e_2 + m_2 )+ g_2 s_2 f_1 (2e_1 + m_1 )\\
& +f_1 f_2(e_1m_2 + e_2 m_1 + 2e_1 e_2 ))+ f_1 f_2(m_1 m_2)\\
= & 2e_{mult} +f_1 f_2(m_1 m_2)
\end{split}
\end{equation}
Here, it is seen that multi-key homomorphic operation increases noise more than a single key homomorphic evaluation. However, $m_1+m_2$ and $m_1 m_2$ can still be recovered correctly using the jointly obtained secret key since $f,g,s,e$ all are sampled from the bounded distribution $\chi$. In other words, the decryption still works if the each of the noise parameters $e_{add}$ and $e_{mult}$ are smaller than $|p/2|$. 
\end{itemize}

\par As observed in all of the FHE schemes presented in detail in our work, since in \cite{lopez2012fly} noise grows with homomorphic operations on encrypted data, the proposed scheme is actually an SWHE scheme. To make it fully homomorphic, L{\'o}pez-Alt et al. also (like all others above) used Gentry's bootstrapping technique. However, to apply bootstrapping, one first needs to  make the underlying SWHE scheme bootstrappable. For this reason, first modulus reduction technique described in \cite{brakerski2012fully,brakerski2014efficient} was used. Then, the final scheme was named a leveled-FHE because it had the ability to deal only a limited number of public keys. Although the number of parties that can be used in homomorphic operations is limited, the complexity of circuit that can be used in homomorphic operations is still independent of the number of parties that  can join the communication. 

\par Another issue to be taken account in \cite{lopez2012fly} is the assumptions. Specifically, two assumptions are used in the scheme proposed by Lopez-Alt et al. First is RLWE problems and second is Decisional Small Polynomial Ratio (DSPR). Though RLWE is well-studied and about being a standard problem, DSPR assumption is a non-standard one. Hence, in~\cite{bos2013improved}, Bos et al. showed how to modify~\cite{lopez2012fly} to remove DSRP assumption. While removing DSRP assumption, the \textit{tensoring} technique introduced in \cite{brakerski2012fully} is used to restrict the noise \revision{increase during homomorphic operations. 
%It is also noted that the new scheme is scale invariant, but not a multi-key scheme.
However, the tensoring technique used to avoid DSRP assumption results in a large evaluation key and a complicated key switching procedure, which makes the scheme impractical. A practical variant of their scheme, which reintroduces the DSRP assumption is also presented in the same work. However, it is later shown that the optimizations and parameter selection that yield a significant increase in the performance makes it vulnerable to sub-field lattice attacks~\cite{albrecht2016subfield}. The attack shown by Albrecht et al. affected not only~\cite{bos2013improved}, but every other NTRU-like scheme, which relies on DSRP problem and whose parameters (e.g., secret key, modulus) are chosen poorly. %The attack works when the secret key f is chosen from a narrow distribution, e.g. ||f|| 5 √q and when the polynomial modulus is chosen such that a subfield of reasonable size exists. In this setting, Albrecht et al. show that the DSPR problem is not as hard as believed thereby invalidating the basic assumption in the LTV [32,17] and YASHE’ [1] schemes. Thus, the subfield lattice attack significantly diminishes the asymptotic security of both schemes.
Finally, in \cite{doroz2016flattening}, a modified NTRU-like FHE scheme, which does not require the DSRP assumption, thereby secure against subfield lattice attacks, is proposed. Another attractive feature of the new FHE scheme is that it also does not require the use of evaluation key during the homomorphic operations. The new scheme is based on~\cite{stehle2011making} and it uses a \textit{Flattening} noise management technique adopted from the flattening technique of~\cite{gentry2013homomorphic}.} 
%Furthermore, in \cite{bos2013improved}, it is also shown how to use larger inputs via Chinese Remainder theorem (CRT). 

\par  Two follow-up interesting works \revision{also} improved the NTRU-like FHE using different techniques. While one of them focuses on a customized and a generic bit-sliced implementation of NTRU-like FHE schemes \cite{doroz2014homomorphic} and the other suggests the use of GPU \cite{dai2014accelerating}. Furthermore, in \cite{doroz2014homomorphic}, the AES circuit is chosen to evaluate the homomorphic operations, which is faster than the proposed one in \cite{gentry2012homomorphic}. Other improvements on hardware implementations of NTRU-like FHE schemes are more recently published in \cite{liu2015efficient,doroz2015accelerating}. Another NTRU-like FHE scheme was suggested in \cite{rohloff2014scalable}. They used the bootstrapping proposed in \cite{alperin2013practical} and "double-CRT" proposed in \cite{gentry2012homomorphic} to modify the representation of the ciphertexts in more efficient way.

\section{Implementations of SWHE and FHE schemes} \label{sec:implementations}
\par The ultimate goal with different HE schemes is to obtain an unbounded and practical FHE scheme. PHE schemes and SWHE schemes proposed before Gentry's breakthrough work in 2009 were stepping stone towards that goal. Nonetheless, they are restricted in terms of the areas that can be applied. However, the SWHE schemes proposed after Gentry's work are mostly the part of the FHE schemes rather than a different scheme. Moreover, a bounded (level) FHE can also be called as SWHE scheme. Hence, it is not possible to separate SWHE and FHE schemes for the works proposed after Gentry's work. In this section, we summarize the implementations of the SWHE and FHE schemes, which can lead to the new works and speed up the follow-up works, proposed after Gentry's work.

\begin{table}[t]
\caption{"Fully" implemented FHE schemes \label{table1}}
\renewcommand{\arraystretch}{1.25}
\resizebox{\textwidth}{!}{
\begin{tabular}{|l|C{3cm}|c|C{1.8cm}|c|c|c|C{1.5cm}|C{1.5cm}|C{1.5cm}|}
\hline
\multicolumn{2}{|c|}{\textbf{Scheme Information}} & \multicolumn{1}{c|}{\textbf{Platform}} &  \multicolumn{2}{c|}{\textbf{Parameters}} & \multicolumn{5}{c|}{\textbf{Running Times}} \\ \hline
\textbf{Implemented Scheme} & \textbf{Base Scheme} & \textbf{Software} & \textbf{Security parameter, $\lambda$} & \textbf{dimension, $n$} & \textbf{PK size} & \textbf{KeyGen} & \textbf{Enc} & \textbf{Dec} & \textbf{Recrypt}  \\ \hline \hline
GH11~\cite{gentry2011implementing} & Gen09~\cite{gentry2009fully}  & C/C++ & 72 & 33768 & 2.25 GB & 2.2 h & 3 min (SWHE) & 0.66 s (SWHE) & 31 min \\ \hline
CMNT11~\cite{coron2011fully} & DGHV10~\cite{van2010fully}  & Sage 4.5.3 & 72 & 7897 & 802 MB & 43 min & 2 min 57 s & 0.05 s & 14 min 33 s \\ \hline
CNT12 (with compressed PK)~\cite{coron2012public} & DGHV10~\cite{van2010fully}  & Sage 4.7.2 & 72 & 7897 & 10.3 MB & 10 min & 7 min 15 s & 0.05 s & 11 min 34 s \\ \hline
CNT12 (leveled)~\cite{coron2012public} & DGHV10~\cite{van2010fully}  & Sage 4.7.2 & 72 & 5700 & 18 MB & 6 min 18 s & 3.4 s & 0.00 s & 2 h 27 min \\ \hline
\end{tabular}
}
\end{table}

\par Implementation of a cryptographic scheme is the middle step between designing the scheme and applying it to a real life service and it provides a realistic performance assessment of the designed scheme. Although some new proposed FHE schemes have increased the efficiency and performance of the implementations significantly, the overhead and cost of the FHE implementations are still too high to be applied transparently in a real life service without disturbing the user.

\subsubsection{\revision{"Fully" implemented FHE schemes}}
After solving the long term open problem of designing a fully homomorphic scheme \cite{gentry2009fully}, many new fully homomorphic scheme proposals were tested with implementation. In a very first attempt, Smart and Vercauteren implemented their scheme in \cite{smart2010fully}, which is a variant of Gentry's original scheme. However, their key generation takes hours up to $N=2^{11}$, where $N$ is the lattice dimension and does not generate the key pairs after $N=2^{11}$. More importantly, their implementation did not include the bootstrapping procedure. Hence, it is actually a SWHE scheme as it was implemented. Then, Craig Gentry and Shai Halevi \cite{gentry2011implementing} succeeded to implement the FHE scheme first time by continuing the way that Smart and Vercauteren had started. The running times for the implementation in [23] and other proposed FHE implementations which are evaluated over random depth circuits are given in Table~\ref{table1}. Moreover, Gentry and Halevi in \cite{gentry2011implementing} introduced some optimizations and simplifications on the squashing process to obtain a bootstrappable scheme. In their implementation, they showed four security levels: toy, small, medium, and large. They suggested that the large parameter settings are practically secure, which have a lattice dimension of $2^{15}$. However, the performance of the implementation is very inefficient in practical terms. For the large parameter setting, a key pair was generated at $2.2$ hours and public key size was $2.25$ GB. Recrypting the ciphertexts (bootstrapping) took $31$ minutes. After that,  in \cite{coron2011fully}, an integer variant of the FHE scheme introduced originally in \cite{van2010fully} was implemented. In this implementation, the key generation takes $43$ min, and the public key size is $802$ MB. The implementation showed that the same security level can be achieved with a much simpler scheme. (The difference comes from the different definitions of security levels). Later, Coron et al. in a different work \cite{coron2012public} improved public key size to $10$ MB, key generation to $10$ minutes, and recryption procedure to 11 min 34 seconds using the similar parameter settings in \cite{coron2011fully}. This performance is obtained using a compression technique on the public key. In \cite{coron2012public}, a leveled DGHV scheme is also implemented with slightly worse performance. Yuanmi Chen and Phong Q. Nguyen~\revision{\cite{chen2012faster}} proposed an algorithm to break the scheme in~\cite{coron2012public}, which is faster than exhaustive search. This work showed that the security level of the scheme proposed in \cite{coron2012public} is much lower than the scheme proposed in \cite{gentry2011implementing}.

\begin{table}[t]
\centering
\caption{FHE implementations for "Low-depth" circuits \label{table2}}
\renewcommand{\arraystretch}{1.25}
\resizebox{\textwidth}{!}{
\begin{tabular}{|l|C{3cm}|C{2cm}|c|c|c|c|c|c|}
\hline
\multicolumn{ 2}{|c|}{\textbf{Scheme Information}} & \textbf{Platform} & \multicolumn{ 2}{c|}{\textbf{Parameters}} & \multicolumn{ 4}{c|}{\textbf{Running times}} \\ \hline
\textbf{Implemented Scheme} & \textbf{Base Scheme} &  \textbf{Software} &   &  & \textbf{Enc} & \textbf{Dec} & \textbf{Mult} & \textbf{Add} \\ \hline \hline
NLV11~\cite{naehrig2011can} & BV11~\cite{brakerski2011fully}  & Magma & $w=2^{32}$  & q=127 & 756 ms & 57 ms & 1590 ms & 4 ms \\ \hline
YASHE (by BLLN13~\cite{bos2013improved}) & LTV12~\cite{lopez2012fly}  & C/C++ & $t=2^{10}$ & q=130 & 27 ms & 5 ms & 31 ms & 0.024 ms \\ \hline
YASHE (by LN14a~\cite{lepoint2014comparison}) & LTV12~\cite{lopez2012fly} & C/C++ & $w=2^{32}$  & q=130 & 16 ms & 15 ms & 18 ms & 0.7 ms \\ \hline
FV (by LN14a~\cite{lepoint2014comparison}) & BV11~\cite{brakerski2011fully} & C/C++ & $w=2^{32}$  & q=130 & 34 ms & 16 ms & 59 ms & 1.4 ms \\ \hline
RC14~\cite{rohloff2014scalable} & LTV12~\cite{lopez2012fly} & Matlab & $n=2^{10}$ & t=1 & 12 ms & 3.36 ms & 100 ms & 0.56 ms \\ \hline
\end{tabular}
}
\end{table}

\subsubsection{\revision{FHE implementation for "Low-depth" circuits}}The second type of FHE implementations tried to implement leveled-FHE schemes for small depth circuits with given run time for isolated and composed addition and multiplication \cite{naehrig2011can,bos2013improved,lepoint2014comparison,rohloff2014scalable}. The comparisons for these small-depth FHE implementations are given at Table~\ref{table2}. Since the performance of the state of the art was unsatisfactory, as an early attempt,  a relatively simpler FHE, which allows only a few homomorphic multiplication operations was implemented in \cite{naehrig2011can}. Later, this performance was improved by Bos et al.~\cite{bos2013improved} due to the new method to evaluate the homomorphic multiplication operation. Moreover, unlike \cite{naehrig2011can}, in \cite{bos2013improved} the underlying scheme was implemented  in C programming language to avoid the unwelcome overhead due to the computer algebra system. Then, a similar performances with~\cite{bos2013improved} is obtained. Recently, a significant improvement is made by using double-CRT in the representation of ciphertexts and used parallelism to accelerate the implementation in Matlab \cite{rohloff2014scalable}.

\subsubsection{\revision{ "Real world" complex FHE implementations}} In contrast to above schemes, which are either proof of concept or small-depth implementations, the authors in \cite{gentry2012homomorphic} implemented FHE for the first time to evaluate \revision{the} circuit complex enough for a real life application. In \cite{gentry2012homomorphic} Gentry et al. implemented a variant of BGV scheme proposed in \cite{cryptoeprint:2011:277}\footnote{\revision{Later updated in~\cite{brakerski2012leveled}}.}, which is a leveled FHE without bootstrapping, in order to evaluate AES circuit homomorphically. Actually, the idea of homomorphic evaluation of AES is first discussed in \cite{naehrig2011can} with the following scenario. A client first sends the key of AES by encrypting with FHE, $FHE(K)$. Then, the client uploads the data by encrypting with AES only, ${AES}_K(m)$. When the cloud wants to evaluate the data homomorphically, it computes $FHE({AES}_K(m))$ and decrypts AES homomorphically (blindfold) to obtain $FHE(m)$. After that, the cloud can compute every homomorphic operation on the data encrypted with FHE. The comparison of such more complex "real world" FHE implementations are presented in Table~\ref{tablereal}. A realization of how to achieve SIMD (single-instruction multiple-data) operations using homomorphic evaluation of AES is proposed by  Smart and Vercauteren \cite{cryptoeprint:2011:133}. Later, some works \cite{cryptoeprint:2013:036,mella2013homomorphic,coron2014scale,doroz2014homomorphic} also improved the performances of the homomorphic evaluation of AES circuit by applying the recent improvements and optimizations in theoretical side. In addition to the use of AES circuit to evaluate homomorphically, lightweight block ciphers such as Prince \cite{doroz2014toward}, SIMON \cite{lepoint2014comparison}, and LowMC \cite{albrecht2015ciphers} are also proposed. In \cite{mella2013homomorphic}, Mella and Susella estimated the cost of some of the symmetric cryptographic primitives such as AES-128, SHA-256 hash function, Salsa20 stream cipher, and KECCAK sponge function. They concluded that AES is best suited for the homomorphic evaluation because of its low number of rounds and absence of integer operations and logical ANDs in its internals. However, in \cite{mella2013homomorphic}, only AES-128 is implemented. 

\subsubsection{\revision{Publicly available FHE implementations}} Although all aforementioned implementations are published in the literature, unfortunately, only a few of them are publicly available to researchers. Some of the publicly available implementations are listed in Table~\ref{tableopen}. From publicly available implementations, HElib \cite{HElib} is the most important and widely utilized one. HElib implements the BGV scheme \cite{cryptoeprint:2011:277} with Smart-Vercauteren ciphertext packing techniques and some new optimizations. The design and implementation of HElib are documented in \cite{halevi2013design} and algorithms used in HElib are documented in \cite{halevi2014algorithms}. HElib is designed using low-level programming, which deals with the hardware constraints and components of the computer without using the functions and commands of a programming language and hence, defined as "assembly language for HE". It was implemented using GPL-licensed C++ library. Since December 2014, it supports bootstrapping \cite{halevi2015bootstrapping} and since March 2015, it supports multi-threading. In an important extension, homomorphic evaluation of AES was implemented on top of HElib \cite{gentry2012homomorphic} and included in the HElib source code in \cite{HElib}.

\begin{table}[t]
\caption{"Real world" complex FHE implementations \label{tablereal}}
\renewcommand{\arraystretch}{1.35}
\resizebox{\textwidth}{!}{
\begin{tabular}{|l|C{5cm}|c|C{2.5cm}|c|C{1cm}|C{1.5cm}|C{1.5cm}|C{1cm}|}
\hline
\multicolumn{ 3}{|c|}{\textbf{Scheme}} & \revision{\textbf{Platform}} & \multicolumn{ 2}{c|}{\textbf{Parameters}} & \multicolumn{3}{c|}{\textbf{Running Times}}  \\ \hline
\multicolumn{1}{|c|}{\textbf{Implemented Scheme}} & \textbf{Base scheme} & \textbf{Circuit} & \revision{\textbf{Reported Specs}}  & \textbf{$\lambda$} & \textbf{AND depth} & \textbf{total evaluation time} & \textbf{number of parallel enc} & \textbf{relative time}  \\ \hline \hline
GHS12 (original)(packed)~\cite{gentry2012homomorphic} & \multicolumn{ 1}{c|}{BGV11~\cite{cryptoeprint:2011:277}} & \multicolumn{ 1}{c|}{AES} & \multicolumn{ 1}{C{2.5cm}|}{ \revision{Intel Xeon CPU @ 2.0 GHz with 256GB RAM}} & \multicolumn{ 1}{c|}{80} & \multicolumn{ 1}{c|}{40}& 48 hours & 54 & 37 min  \\ \cline{ 1- 1}\cline{ 7- 9}
GHS12 (original)(byte-sliced)~\cite{gentry2012homomorphic} & \multicolumn{ 1}{l|}{} & \multicolumn{ 1}{c|}{} & \multicolumn{ 1}{c|}{} & \multicolumn{ 1}{c|}{}& \multicolumn{ 1}{l|}{} & 65 hours & 720 & 5 min  \\ \hline
CLT13 (byte-wise)~\cite{cheon2013batch} & \multicolumn{ 1}{c|}{DGHV10~\cite{van2010fully}} & \multicolumn{ 1}{c|}{AES} & \multicolumn{ 1}{C{2.5cm}|}{ \revision{Intel Core i7 @ 3.4Ghz with 32GB RAM}} & \multicolumn{ 1}{c|}{72} & \multicolumn{ 1}{c|}{40} & 18.3 hours & 33 & 33 min  \\ \cline{ 1- 1}\cline{ 7- 9}
CLT13 (state-wise)~\cite{cheon2013batch} & \multicolumn{ 1}{l|}{} & \multicolumn{ 1}{l|}{} & \multicolumn{ 1}{c|}{} & \multicolumn{ 1}{c|}{} & \multicolumn{ 1}{c|}{} & 113 hours & 531 & 12 min 46 s  \\ \hline
CLT14 (state-wise)~\cite{coron2014scale} & \multicolumn{ 1}{c|}{DGHV10~\cite{van2010fully}} & \multicolumn{ 1}{c|}{AES} & \multicolumn{ 1}{C{2.5cm}|}{ \revision{Intel Xeon E5-2690 @ 2.9 GHz}} & 80 & \multicolumn{ 1}{c|}{40} & 102 hours & 1875 & 3 min 15 s \\ \cline{ 1- 1}\cline{5- 5}\cline{ 7- 9}
CLT14 (state-wise)~\cite{coron2014scale} & \multicolumn{ 1}{l|}{} & \multicolumn{ 1}{l|}{} & \multicolumn{ 1}{c|}{} & 72 & \multicolumn{ 1}{c|}{} & 3 h 35 min & 569 & 23 s  \\ \hline
LN14a (YASHE)~\cite{lepoint2014comparison} & LTV12~\cite{lopez2012fly} & \multicolumn{ 1}{c|}{SIMON} & \multicolumn{ 1}{C{2.5cm}|}{ \revision{Intel Core i7-2600 @ 3.4 GHz\tablefootnote{\revision{With hyper-threading turned off and over-clocking (‘turbo boost’) disabled.}}}} & \multicolumn{ 1}{c|}{128} & \multicolumn{ 1}{c|}{34} & 1 h 10 min & 2048 & 2.04 s  \\ \cline{ 1- 1} \cline{ 2- 2}\cline{ 7- 9}
LN14a (FV)~\cite{lepoint2014comparison} & Bra12~\cite{brakerski2012fully} & \multicolumn{ 1}{c|}{}  & \multicolumn{ 1}{c|}{} & \multicolumn{ 1}{c|}{} & \multicolumn{ 1}{c|}{} & 3 h 27 min & 2048 & 6.06 s \\ \hline
DHS14~\cite{doroz2014homomorphic} & \multicolumn{ 1}{c|}{LTV12~\cite{lopez2012fly}} & AES &  { \revision{Intel Xeon @ 2.9 GHz}} & $\sim$80 & 40 & 31 hours & 2048 & 55 s  \\ \hline
DSES14~\cite{doroz2014toward} & \multicolumn{ 1}{c|}{LTV12~\cite{lopez2012fly}} & Prince &{ \revision{Intel Core i7 3770K @
3.5 Ghz with 32 GB RAM\tablefootnote{\revision{Only single thread is used.}}}} & 130 & 30 & 57 min & 1024 & 3.3 s  \\ \hline
ARSTZ15~\cite{albrecht2015ciphers} & BGV11~\cite{cryptoeprint:2011:277}  & LowMC  & {\footnotesize \revision{Intel Haswell i7-4770K CPU @ 3.5 GHz with 16GB RAM}} & 80 & 12 & 8 min & 600 & 0.8 s \\ \hline
GHS12 (updated)(no bootstrapping)~\cite{gentry2012homomorphic} & \multicolumn{ 1}{c|}{BGV11~\cite{cryptoeprint:2011:277}} & \multicolumn{ 1}{c|}{AES } & \multicolumn{ 1}{C{2.5cm}|}{ \revision{Intel Core i5-3320M at 2.6GHz with 4GB RAM\tablefootnote{\revision{An Ubuntu 14.04 installed VM}}}} & \multicolumn{ 1}{c|}{80} & \multicolumn{ 1}{c|}{40} & 4 min 12 s & 120 & 2 s  \\ \cline{ 1- 1}\cline{ 7- 9}
GHS12 (updated)(with bootstrapping)~\cite{gentry2012homomorphic} & \multicolumn{ 1}{l|}{} & \multicolumn{ 1}{l|}{} & \multicolumn{ 1}{c|}{} & \multicolumn{ 1}{c|}{} & \multicolumn{ 1}{c|}{} & 17 min 30 s & 180 & 5.8 s  \\ \hline
\end{tabular}
}
\end{table}

\begin{table}[t]
\centering
\caption{Some publicly available FHE implementations \label{tableopen}}
\renewcommand{\arraystretch}{1.4}
\resizebox{\textwidth}{!}{
\begin{tabular}{|c|c|C{1cm}|c|c|}
\hline
\textbf{Name} & \textbf{Scheme} & \textbf{\revision{Lang}} & \textbf{Documentation} &  \textbf{Libraries}\\ \hline	
\specialcell{HElib\\ \cite{HElib}} & \specialcell{BGV\\ \cite{cryptoeprint:2011:277}}  & C++ & \specialcell{Yes\\ \cite{halevi2013design}} & NTL, GMP \\ \hline
\specialcell{libScarab\\ \cite{libScarab}} & \specialcell{SV\\ \cite{smart2010fully}} & C & \specialcell{Yes\\ \cite{perl2011poster}} &  \specialcell{GMP, FLINT,\\MPFR, MPIR} \\ \hline
\specialcell{FHEW\\ \cite{FHEW}} & \specialcell{DM14\\ \cite{ducas2015fhew}} & C++ & \specialcell{Yes\\ \cite{ducas2015fhew}} &  FFTW \\ \hline
\specialcell{\revision{TFHE}\\ \revision{\cite{TFHE}}} & \specialcell{\revision{CGGI16}\\ \revision{\cite{chillotti2016faster}}} & \revision{C++} & \specialcell{\revision{Yes}\\ \revision{\cite{chillotti2016faster}}} &  \revision{FFTW} \\ \hline
\specialcell{\revision{SEAL}\\ \revision{\cite{SEAL}}} & \specialcell{\revision{FV12}\\ \revision{\cite{cryptoeprint:2012:144}}} & \revision{C++} & \specialcell{\revision{Yes}\\ \revision{\cite{SEAL_doc}}} &  \specialcell{\revision{No external} \\ \revision{dependency}} \\ \hline

\end{tabular}
}
\end{table}

\par Unfortunately, the usage of HElib is not easy because of the sophistication needed for its low-level implementation and parameter selection which effects both performance and security level. Another notable open source FHE implementation is libScarab \cite{libScarab}. To the best of our knowledge, libScarab \cite{libScarab} is the first open-source implementation of FHE. Its parameter selection is relatively easier than that of HElib, but it suffers from a lot of limitations. For instance, it does not implement modern techniques (e.g., modulus reduction and re-linearization techniques \cite{brakerski2014efficient}) to handle the noise level or it also does not support the SIMD techniques introduced in \cite{smart2014fully}.  It implements Smart-Vercauteren's FHE scheme in \cite{smart2010fully} and documentation is provided in \cite{perl2011poster}. 

\par Another major implementation is introduced by Ducas and Micciancio and called "Fastest Homomorphic Encryption in the West" (FHEW) \cite{FHEW}. It is documented in \cite{ducas2015fhew}. It significantly improves the time required to bootstrap the ciphertext claiming homomorphic evaluation of a NAND gate "in less than a second". A NAND gate is functionally complete. Hence, any possible boolean circuits can be built using only NAND gates. In \cite{ducas2015fhew}, the usage of ciphertext packing and SIMD techniques provides an amortized cost. However, in FHEW such performance is achieved using only a few hundred lines of code with the use of one additional library, FFTW \cite{FFTW05}. \revision{Later, the homomorphic computation cost of any binary gate~\cite{ducas2015fhew} is increased by a factor of 50 by making some optimizations on the bootstrapping algorithm. The main improvement is based on the torus representation of LWE ciphertexts. This improved the cost of bootstrapping 10 times according to the best known bootstrapping in~\cite{FHEW}. They also further improved the noise propagation overhead algorithms using some approximations. Finally, they also reduced the size of bootstrapping key from 1GB to 24MB by achieving the same security level.}

\revision{More recently, another HE library called Simple Encrypted Arithmetic Library (SEAL)~\cite{SEAL} is released by Microsoft. The goal of releasing this library is explained as providing a well-documented HE library that can be easily used by both crypto experts and non-experts with no crypto background like practitioners in bioinformatics. The library does not have external dependencies like others and it includes automatic parameter selection and noise estimator tools, which makes it easier to use. Finally, the security estimates of two well-known LWE-based HE libraries, HElib and SEAL, against dual lattice attacks are revised in~\cite{albrecht2017dual}. It is shown that the parameters promising 80 bits of security actually gives an estimated cost of 68 bits for SEAL v2.0 and 62 bits for HElib. As a final note, we give the list of general-purpose HE libraries as follows: HEAAN implementing that supports fixed point arithmetics~\cite{HEAAN}, a GPU-accelerated library cuHE~\cite{cuHE}, a general lattice crypto library PALISADE~\cite{PALISADE}.}

\subsubsection{\revision{FHE hardware implementations and productions}} The first known usage of FHE in a production environment is announced by Fujitsu Laboratories Ltd. \cite{Fujitsu}. Their reported implementation provides statistical calculations and biometric authentication by using FHE-based security. They improved an FHE by batching the string bits of data. The practical testing of this FHE implementation by Fujitsu is still pending as of this writing. Although the software only implementations are considered promising to obtain a practical FHE implementation, there is still a substantial gap between the achieved and the targeted performance. This gap led to new alternative research area in hardware implementations. The hardware solutions to accelerate both FHE and SWHE schemes mainly focused on three implementation platforms: Graphics Processing Unit (GPU), Application-Specific Integrated Circuit (ASIC), and Field-Programmable Gate Array (FPGA) (A useful survey of hardware implementations of homomorphic schemes can be found in \cite{moore2014practical}). Although GPU is for graphical purposes, its highly parallel structure offers great promise over CPU for efficiency. Hence, it is suggested in some studies to use GPU order to improve the efficiency of homomorphic evaluation \cite{dai2014accelerating,wang2014accelerating,wang2015exploring,daiaccelerating,lee2015accelerating}. One of the major barriers to a practical FHE is the noise growth in the homomorphic multiplication operation. This prompted researchers to find a solution that can deal with a large number of modular multiplications. Therefore, there are some works focusing particularly on this problem using the customized ASICs \cite{doroz2013evaluating,wang2014vlsi,doroz2013accelerating}. In spite of the potential of GPU and ASIC solutions, most of the proposed studies are based on the reconfigurable hardware, specifically FPGA. FPGA platforms offer not only Fast Fourier Transform (FFT), but also some optimization techniques such as number theoretic transformation (NTT) and fast modular polynomial reduction at hardware level. Such large and reconfigurable environment provided by FPGAs motivates many researchers to speed up the practicality of FHE schemes \cite{cousins2012update,wang2013fpga,cao2013accelerating,moore2013targeting,chen2015high,cao2014high,moore2014accelerating,cousins2014fpga,roymodular,poppelmann2015accelerating,ozturk2015accelerating}.

\par In conclusion, some of the SWHE implementations (leveled-FHE) \cite{gentry2012homomorphic} 
get closer to a tolerable performance. However, the bootstrapping techniques in FHE schemes need to be improved and the cost of homomorphic multiplications should be reduced to increase the performance.
\section{Further Research Directions and Lessons Learned} \label{sec:further}
Performance of any encryption scheme is evaluated with three different criteria: security, speed, and simplicity. First, an encryption scheme must be secure so that an attacker can not obtain any type of information by using a reasonable amount of resources. Second, its efficiency must not disturb the user's comfort, i.e., it must be transparent to the users because users prefer usability against security. Lastly, if and only if an encryption scheme is understandable by the other area practitioners, they will implement the scheme for their applications and productions. If the existing FHE schemes are evaluated in terms of the three criteria, there is, though getting closer, still a substantial room for improvement in terms of all these criteria, especially for the speed performance.

\par Even though some of the nonstandard security assumptions (e.g., SSSP\footnote{Indeed, Moon Sung Lee showed that it is quite probable that SSSP challenges can be solved within two days \cite{lee2011sparse,SSSP}.} \cite{lee2011sparse,SSSP}) in the Gentry's original scheme are later removed, there are still some open security issues about the FHE schemes. First one is the circular security of FHE. Circular security (aka KDM security), as mentioned earlier, keeps its own secret key secure by encrypting it with the public key. All known FHE schemes use Gentry's blueprint bootstrapping technique to obtain an unlimited FHE scheme. So, the encryption of the secret key is also sent to the cloud to bootstrap the noisy ciphertexts and an eavesdropper can capture the encryption of secret key. Even though some SWHE and leveled-FHE schemes are proven as semantically secure, an unbounded FHE still has not been proven as semantically secure with respect to any function, so it does not guarantee that an adversary can not reveal the secret key from its encryption under the public key. This unfortunate situation is still open to be proven. Moreover, although some SWHE schemes \cite{loftus2011cca} are proven as indistinguishable under non-adaptive chosen ciphertext attack (IND-CCA1), none of the unbounded FHE schemes is IND-CCA1 secure for now. (IND-CCA2 (adaptive) is not applicable to FHE because FHE itself requires to be malleable.) In brief, FHE still needs to be studied extensively to prove that it is secure enough.

\par FHE allows an unlimited number of functions on encrypted data. However, limitations on the efficiency of the FHE schemes prompts researchers to find the SWHE schemes that can be good enough to use in real-life applications. Recently, homomorphic evaluation of one AES, which is a highly complex and nontrivial function, is reduced to 2 seconds \cite{gentry2012homomorphic} and researchers are now focusing to improve this instead of trying to implement an FHE scheme, which is extremely slow for now.

\par The main process that increases the computational cost in FHE is the bootstrapping process. An unbounded FHE scheme that allows unlimited operations without bootstrapping is still an open problem. Indeed, the bootstrapping is necessary to decrease the noise in the evaluated ciphertexts. Hence, though a framework was suggested in \cite{nuida2014simple}, the design of noise-free FHE scheme is also one of the open problems. A noise-free FHE \cite{liu2015practical} and an FHE without bootstrapping \cite{yagisawa2015fully} are reported as insecure in \cite{wangnotes}.

\par Showing the existence of FHE instilled hope to solve other long waiting problems (applications) such as Functional Encryption (FE) (i.e., Identity-based encryption (IBE) and Attribute-based encryption (ABE)). Functional encryption basically controls the access over data while allowing computation on it according to the features of identity or attribute. The purpose of designing ABE or IBE based on FHE is to take the advantage of the functionality of two worlds. However, for now, there exists a few \cite{gentry2013homomorphic,clear2014bootstrappable,clearattribute,wang2015efficient}. Another fruitful application of FHE is multi-party computation (MPC) which allows the computation of the function with multiple inputs from different users while keeping the inputs hidden. Even though there exist a few FHE-based MPC protocols \cite{damgaard2012multiparty,lopez2012fly,choudhury2013between,damgaard2016adaptively} proposing these powerful and useful tools, unfortunately, their performances are not yet comparable with the conventional MPC approaches \cite{mood2014reuse,carter2013secure,premnath2014practical,carter2015outsourcing} because of the computational cost of the existing FHE schemes. However, FHE does not require any interaction, which reduces the complexity of the communication protocol significantly. However, there are still some gaps on how to realize those protocols. Furthermore, FHE itself can not perform a homomorphic evaluation on independently encrypted data, i.e., multi-key FHE. some primitive result to deal with this issue was presented in \cite{lopez2012fly}. However, the proposed scheme can only handle a bounded number of users.  When the cloud and number of connected devices are considered, the restriction may not be feasible. Hence, a multi-key FHE with an unlimited number of users is another promising direction for future applications. %will be used in
%To conclude, FHE is a

\section{Conclusion}
In today’s always-on, Internet-centric world, the privacy of data plays a more significant role than ever before. For highly sensitive systems such as online retail and e-banking, it is crucial to protect users’ accounts and assets from malicious third parties. Nonetheless, today's norm is to encrypt the data and share the keys with the service provider, cloud operator, etc. In this model, the control over the privacy of the sensitive data is lost. The users or service providers with the key have exclusive rights on the data. Untrusted providers, cloud operators can keep sensitive data and its identifying credentials of users long after the user ends the relationship with the services. One promising direction to preserve the privacy of the data is to utilize homomorphic encryption (HE) schemes. HE is a special kind of encryption scheme, which allows any third party to operate on the encrypted data without decrypting it in advance. Indeed, the idea of HE has been around for over 30 years; however, the first plausible and achievable \textit{Fully Homomorphic Encryption} (FHE) scheme was introduced by Craig Gentry in 2009. Since then, different FHE schemes demonstrated that FHE still needs to be improved significantly to be practical on every platform as they are very expensive for real-life applications. Hence, in this paper, we surveyed the HE and FHE schemes. Specifically, starting from the basics of HE, the details of the well-known \textit{Partially HE} (PHE) and \textit{Somewhat HE} (SWHE), which are important pillars of achieving FHE, were presented. Then, after classifying FHE schemes in the literature under four different categories, we presented the major FHE schemes with this classification. Moreover, we articulated the implementations and the new improvements in Gentry-type FHE schemes. Finally, we discussed promising research directions as well as lessons learned for interested researchers. 

\bibliographystyle{ACM-Reference-Format-Journals}
\bibliography{ref-acmsmall-sample-bibfile}
                             % Sample .bib file with references that match those in
                             % the 'Specifications Document (V1.5)' as well containing
                             % 'legacy' bibs and bibs with 'alternate codings'.
                             % Gerry Murray - March 2012

% History dates
%\received{February 2007}{March 2009}{June 2009}

\end{document}